\documentclass[aip,jcp,unsortedaddress,12pt,preprint]{revtex4-1}

\usepackage{amsthm,amsfonts,amssymb,amsmath,dsfont}
\usepackage[T1]{fontenc}
\usepackage{graphicx,graphics,color}
\usepackage{colortbl}

\usepackage{natbib}

\usepackage[colorlinks=true,linkcolor=blue]{hyperref}

\begin{document}

\newcommand{\nn}{\nonumber}
\newcommand{\Tr}{\mathrm{Tr}}
\newcommand{\Ln}{\mathrm{Ln}}
\newcommand{\bra}{\langle}
\newcommand{\ket}{\rangle}
\newcommand{\del}{\partial}
\newcommand{\vt}{\vec}
\newcommand{\dg}{^{\dag}}
\newcommand{\cg}{^{*}}
\newcommand{\vep}{\varepsilon}
\newcommand{\suml}{\sum\limits}
\newcommand{\prodl}{\prod\limits}
\newcommand{\intl}{\int\limits}
\newcommand{\til}{\widetilde}
\newcommand{\mcl}{\mathcal}
\newcommand{\mfk}{\mathfrak}
\newcommand{\ds}{\displaystyle}
\renewcommand{\dim}{\mathrm{dim}}
\renewcommand{\Re}{\mathrm{Re}}
\renewcommand{\Im}{\mathrm{Im}}
\renewcommand{\mod}{\mathrm{\,mod\,}}
\renewcommand{\b}{\overline}

\title{Initial value representation for the $\mathrm{SU}(n)$ semiclassical propagator}

\author{Thiago F. Viscondi}
\email{viscondi@ifi.unicamp.br}
\author{Marcus A. M. de Aguiar}
\affiliation{Instituto de Física `Gleb Wataghin', Universidade Estadual de Campinas, 13083-859, Campinas, SP, Brazil}

\date{\today}

\begin{abstract}
The semiclassical propagator in the representation of $\mathrm{SU}(n)$ coherent states is characterized by isolated classical
trajectories subjected to boundary conditions in a doubled phase space. In this paper we recast this expression in terms of an
integral over a set of initial-valued trajectories. These trajectories are monitored by a filter that collects only the
appropriate contributions to the semiclassical approximation. This framework is suitable for the study of bosonic dynamics in $n$
modes with fixed total number of particles. We exemplify the method for a Bose-Einstein condensate trapped in a triple-well
potential, providing a detailed discussion on the accuracy and efficiency of the procedure.
\end{abstract}



\maketitle

\section{Introduction}
\label{sc:intro}

Semiclassical methods have proved to be very useful in the investigation of systems with many degrees of freedom, especially in
atomic and molecular dynamics\cite{Miller01, Thoss04, Kay05, Miller06}. Moreover, the semiclassical approximation has also been
an important theoretical tool in studying the connection between the classical and quantum theories, particularly in fundamental
topics such as chaos and open quantum systems\cite{Koch08, Moix08, Goletz10}.

The semiclassical propagator in the coordinate representation was first derived by Van Vleck\cite{VanVleck28} at the beginning of
the last century. However, this fundamental result has two remarkable characteristics that considerably hinder its practical
application. First, the Van Vleck propagator is determined by classical trajectories subject to \textit{boundary conditions}. In
general, the search for these specific solutions is quite complicated, particularly in multidimensional and chaotic systems. The
second major problem is the appearance of \textit{focal points}, which are responsible for divergences in the semiclassical
approximation.

A different line of research, concerned with the difficulties caused by focal points, led to the development of semiclassical
propagators in the representation of the harmonic-oscillator coherent states\cite{Klauder85, Baranger01, Fierro07, Braun07}.
Although it has been found that the focal points still persisted, this alternative approach has demonstrated some evident
advantages over the coordinate and momentum representations, including an immediate visualization of the system over the full
phase space. Nevertheless, new problems have emerged, such as the \textit{duplication of the phase space}, resulting from the
apparent overdetermination of the classical equations of motion. Furthermore, not all classical trajectories in the extended
phase space, while correctly satisfying the boundary conditions, correspond to semiclassical propagators with physical
meaning\cite{Huber87, Huber88, Adachi89, Rubin95, Shudo95, Shudo96, Ribeiro04, Aguiar05}. Therefore, it is necessary to establish
effective rules for selecting the proper contributions to the semiclassical dynamics.

In the last decades, many different techniques have been proposed in order to solve the recurrent problems in semiclassical
propagation\cite{Miller70, Miller74, Heller75, Herman84, Kay94a, Kay94b, Kay97, Zhang03, Zhang04, Heller91, Tomsovic91,
Shalashilin04, Shalashilin08, Pollak03, Kay06}. Most of these methods are based on the concept of \textit{initial value
representation}, in which the system dynamics is determined only by initial conditions, avoiding the search for boundary-valued
trajectories.

Recently, Aguiar et al. presented a new approach to the semiclassical propagator of the harmonic-oscillator coherent states,
which combines the unique resources offered by the trajectories in a doubled phase space with the plain advantages of an initial
value representation\cite{Aguiar10}. Moreover, they demonstrated that practical and simple rules for selecting the contributing
trajectories can produce very accurate results.

The procedures developed in the present paper are similar to those of Aguiar et al., but generalized to a subclass of the
$\mathrm{SU}(n)$ coherent states. These states constitute an ideal setting to study the bosonic dynamics for a fixed total number
of particles in $n$ modes. In this paper we also propose a new prescription for the selection of contributing trajectories, which
we designate as a \textit{heuristic filter}. For a detailed derivation of the $\mathrm{SU}(n)$ semiclassical propagator we refer
the reader to a recent work of the present authors\cite{Viscondi11b}.

The remainder of the paper is organized as follows: in section \ref{sc:sclpropmeth} we develop the semiclassical propagation
method based on an initial value representation. We start with a brief review of the $\mathrm{SU}(n)$ coherent states, in which
we introduce fundamental aspects of the adopted notation. Then, we present the $\mathrm{SU}(n)$ semiclassical propagator,
followed by other important definitions, such as the effective classical Hamiltonian, the classical equations of motion and the
doubled phase space. Next we reformulate the semiclassical approximation in terms of a set of initial conditions and a heuristic
filter of trajectories. At the end of the section, we describe the procedure used for calculating semiclassical mean values of
observables, based on the phase space representation of states. Section \ref{sc:appldisc} presents an application of the
$\mathrm{SU}(2)$ and $\mathrm{SU}(3)$ semiclassical propagators. As an example, we consider a simplified model for the dynamics
of a Bose-Einstein condensate in a triple-well potential. In this context, we introduce the classical approximation, which
provides a reference for comparison with the semiclassical results. Also, we discuss the accuracy of the semiclassical
propagation in nonlinear and predominantly linear dynamical regimes, by contrasting the approximations with exact quantum
calculations. Finally, in section \ref{sc:conclu} we present our concluding remarks.

\section{Semiclassical propagation method for $\mathrm{SU}(n)$}
\label{sc:sclpropmeth}

\subsection{$\mathrm{SU}(n)$ coherent states}
\label{ssc:cohstat}

The coherent state related to the fully symmetric irreducible representation of $\mathrm{SU}(n)$ for $N$ identical bosons is
given by\cite{Gilmore75}:

\begin{equation}
  |w\ket=\suml_{m_{1}+m_{2}+\ldots+m_{n}=N}
  \left(\frac{N!}{m_{1}!m_{2}!\ldots m_{n}!}\right)^{\frac{1}{2}}
  \left(\prodl_{j=1}^{n-1}w_{j}^{m_{j}}\right)
  \frac{|m_{1},m_{2},\ldots,m_{n}\ket}{(1+w\cg w)^{\frac{N}{2}}};
  \label{eq1p1}
\end{equation}

\noindent where $\{|m_{1},m_{2},\ldots,m_{n}\ket\}$ is the usual basis of the bosonic Fock space $\mathds{B}^{n}_{N}$ for $n$
modes and $N$ particles, such that $m_{j}$ is the occupation in the $j$-th mode.  The vector
$w=(w_{1},w_{2},\ldots,w_{n-1})^{T}$, with $(n-1)$ complex entries, parametrizes the entire set of coherent states.

Although normalized, the coherent states in \eqref{eq1p1} are not orthogonal\footnote{According to the adopted notation, the
juxtaposition of two vectors $a$ and $b$ represents the matrix product $ab=a_{1}b_{1}+a_{2}b_{2}+\ldots+a_{n-1}b_{n-1}$.}:
\begin{equation}
  \bra w'|w\ket=\frac{(1+w'^{*} w)^{N}}{(1+w'^{*} w')^{\frac{N}{2}}(1+w\cg w)^{\frac{N}{2}}}.
  \label{eq1p2}
\end{equation}

However, due to the overcompleteness of the coherent states, we can write the following diagonal resolution for the identity in
$\mathds{B}_{N}^{n}$:
\begin{equation}
  \intl_{w\,\in\,\mathds{C}^{n-1}} d\mu(w\cg,w)\;|w\ket\bra w|=\mathds{1};\qquad
  d\mu(w\cg,w)=\frac{\sigma(n)\,\dim(\mathbb{B}_{N}^{n})}{(1+w\cg w)^{n}}\prodl_{j=1}^{n-1}d^{2}w_{j};
  \label{eq1p3}
\end{equation}

\noindent where $d^{2}w_{j}=dx_{j}dy_{j}$, with $x_{j}=\Re\left(w_{j}\right)$ and $y_{j}=\Im\left(w_{j}\right)$. Note that the
normalization factor in \eqref{eq1p3} can be divided into $\sigma(n)=\frac{(n-1)!}{\pi^{n-1}}$, which is independent of the total
boson number, and $\dim(\mathbb{B}_{N}^{n}) = \frac{(N+n-1)!}{N!(n-1)!}$, the dimension of the accessible Hilbert space.

\subsection{$\mathrm{SU}(n)$ semiclassical propagator}
\label{ssc:sclpropag}

The quantum propagator in the $\mathrm{SU}(n)$ coherent state representation is defined as the transition probability between the
initial coherent state $|w_{i}\ket$ and final coherent state $|w_{f}\ket$ after a time interval $\tau$\footnote{For simplicity,
in what follows we choose the system of units so that $\hbar=1$.}:
\begin{equation}
  K(w\cg_{f},w_{i};\tau)=\bra w_{f}|e^{-iH\tau}|w_{i}\ket.
  \label{eq1p4}
\end{equation}

After recasting the above propagator as a path integral, we can perform its \textit{semiclassical approximation}, which consists
in expanding the action functional to second order around a classical trajectory. The result of this derivation\cite{Viscondi11b}
is given by\footnote{Considering two vector quantities $a$ and $b$, we denote by $\frac{\del a}{\del b}$ the matrix whose
elements follow from $\left[\frac{\del a}{\del b}\right]_{jk}=\frac {\del a_{j}}{\del b_{k}}$, with $j,k=1,2,\ldots,(n-1)$. In
the case of a scalar function $f(a)$, we have that $\frac{\del f(a)}{\del a}$ represents a vector whose entries are given by
$\left[\frac{\del f(a)}{\del a}\right]_{j}=\frac{\del f(a)}{\del a_{j}}$, also for $j=1,2,\ldots,(n-1)$.}:
\begin{equation}
  K_{sc}(w\cg_{f},w_{i};\tau)=
  e^{i(S+I)-\frac{N}{2}\Ln\left[(1+|w\cg_{f}|^{2})(1+|w_{i}|^{2})\right]}
  \sqrt{\left[\frac{1+\b{w}(\tau)w(\tau)}{1+\b{w}(0)w(0)}\right]^{\frac{n}{2}}
  \det\left[\frac{\del\b{w}(0)}{\del\b{w}(\tau)}\right]}.
  \label{eq1p5}
\end{equation}

All elements of this semiclassical formula are calculated on a classical trajectory, which is solution of the equations of
motion\footnote{In the equation \eqref{eq1p6} we introduce the notation for the dyadic product. That is, considering two
arbitrary vectors $a$ and $b$ of dimension $(n-1)$, the outcome of the product $a\otimes b$ is a matrix with elements given by
$(a\otimes b)_{jk}=a_{j}b_{k}$.}
\begin{equation}
  \left\{
  \begin{aligned}
  \dot{w}&=-\frac{i}{N}(1+\b{w}w)\left[\mathds{1}+w\otimes\b{w}\right]\frac{\del\mcl{H}}{\del\b{w}}
  =-i\xi\frac{\del\mcl{H}}{\del\b{w}}\\
  \dot{\b{w}}&=\frac{i}{N}(1+\b{w}w)\left[\mathds{1}+\b{w}\otimes w\right]\frac{\del\mcl{H}}{\del w}
  =i\b{\xi}\frac{\del\mcl{H}}{\del w}
  \end{aligned}
  \right.
  \label{eq1p6}
\end{equation}
\noindent with boundary conditions
\begin{equation}
  \begin{aligned}
  w(0)&=w_{i};\\
  \b{w}(\tau)&=w\cg_{f}.
  \end{aligned}
  \label{eq1p8}
\end{equation}

In equation \eqref{eq1p6}, $\mathcal{H}$ is the effective classical Hamiltonian:
\begin{equation}
  \mathcal{H}(\b{w},w)=\frac{\bra\b{w}\cg|H|w\ket}{\bra\b{w}\cg|w\ket}.
  \label{eq1p7}
\end{equation}

If the classical equations of motion have more than one solution subject to the same boundary conditions $w_{i}$ and $w_{f}\cg$
with fixed time interval $\tau$, then the correct semiclassical propagator between these points is given by the sum of the
propagators \eqref{eq1p5} for each possible trajectory.

Note that the complex vector variables $w$ and $\b{w}$ are \textit{completely independent}, i.e. in general $\b{w}(t)\neq
w\cg(t)$. This \textit{doubled phase space} is a direct consequence of the introduction of boundary conditions to the equations
of motion. If $\b{w}(t)$ were equal to $w\cg(t)$, the two vector differential equations in \eqref{eq1p6} would be redundant and
the boundary conditions $w(0)=w_{i}$ and $w\cg(\tau)= w\cg_{f}$ would make the problem \textit{overdetermined}. Therefore, the
duplication of the phase space is required to solve the classical equations of motion in the coherent state representation.

The equations of motion \eqref{eq1p6} are derived by the extremization of the following action functional:
\begin{equation}
  \begin{aligned}
  &S(w\cg_{f},w_{i};\tau)=
  \intl_{0}^{\tau}L\left(\b{w},w,\dot{\b{w}},\dot{w}\right)dt
  +\Gamma(w\cg_{f},w_{i};\tau);\\
  &L\left(\b{w},w,\dot{\b{w}},\dot{w}\right)=
  i\frac{N}{2}\frac{\b{w}\dot{w}-\dot{\b{w}}w}{1+\b{w}w}-\mcl{H}(\b{w},w);\\
  &\Gamma(w\cg_{f},w_{i};\tau)=
  -i\frac{N}{2}\Ln\left\{[1+w\cg_{f}w(\tau)][1+\b{w}(0)w_{i}]\right\}.
  \end{aligned}
  \label{eq1p9}
\end{equation}

The function $\Gamma$, known as the boundary term, is essential in obtaining the classical equation of motion subject to the
boundary conditions \eqref{eq1p8}. Another quantity introduced in \eqref{eq1p5} is the correction term to the action\footnote{Due
to the overcompleteness of the coherent states, there are several ways to perform the semiclassical approximation of the
propagator, resulting from different quantization schemes (choices of operator ordering). Each one of these corresponds to a
distinct correction term\cite{Baranger01, Santos06}.}:
\begin{equation}
  I=\frac{1}{4}\intl_{0}^{\tau}\Tr\left[
  \frac{\del}{\del\b{w}}\left(\b{\xi}\frac{\del\mcl{H}}{\del w}\right)
+\frac{\del}{\del w}\left(\xi\frac{\del\mcl{H}}{\del\b{w}}\right)\right]dt
  \label{eq1p10}
\end{equation}
\noindent where the matrices $\xi$ and $\b{\xi}$ are defined in equations \eqref{eq1p6}.

The last ingredient required in the formula \eqref{eq1p5} is the tangent matrix $\mathbb{M}$, governing the dynamics of small
displacements around the classical trajectory, defined in block form by
\begin{equation}
  \left(\begin{array}{c} \delta w(\tau) \\ \delta\b{w}(\tau) \end{array}\right)
  =\left(\begin{array}{c c} M_{11}(\tau) & M_{12}(\tau) \\ M_{21}(\tau) & M_{22}(\tau)
  \end{array}\right)\left(\begin{array}{c} \delta w(0) \\ \delta\b{w}(0) \end{array}\right)
  =\mathbb{M}\left(\begin{array}{c} \delta w(0) \\ \delta\b{w}(0) \end{array}\right).
  \label{eq1p11}
\end{equation}

Notice that
\begin{equation}
  M_{22}(\tau)=\frac{\del\b{w}(\tau)}{\del\b{w}(0)}=
  \left[\frac{\del\b{w}(0)}{\del\b{w}(\tau)}\right]^{-1}
  \label{eq1p12}
\end{equation}
\noindent and,therefore, the block $M_{22}$ is the inverse of the matrix whose determinant appears in the semiclassical
propagator. A focal point in the variables $\b{w}$\footnote{A focal point represents a crossing between trajectories when
projected onto a particular subspace of the complete phase space.} corresponds to a zero value of $\det M_{22}(\tau)$ and,
consequently, to a divergence in \eqref{eq1p5}.

The tangent matrix can be calculated as solution of a system of differential equations subjected to initial conditions. Using
\eqref{eq1p6}, we obtain
\begin{equation}
  \left(\begin{array}{c}
  \delta\dot{w} \\ \delta\dot{\b{w}}
  \end{array}\right)=
  \left(\begin{array}{c c}
  -i\frac{\del}{\del w}\left[\xi\frac{\del\mcl{H}}{\del\b{w}}\right] &
  -i\frac{\del}{\del\b{w}}\left[\xi\frac{\del\mcl{H}}{\del\b{w}}\right]\\
  i\frac{\del}{\del w}\left[\,\b{\xi}\frac{\del\mcl{H}}{\del w}\right] &
  i\frac{\del}{\del\b{w}}\left[\,\b{\xi}\frac{\del\mcl{H}}{\del w}\right]
  \end{array}\right)
  \left(\begin{array}{c}
  \delta w \\ \delta\b{w}
  \end{array}\right)
  =\mathbb{R}
  \left(\begin{array}{c}
  \delta w \\ \delta\b{w}
  \end{array}\right).
  \label{eq1p13}
\end{equation}

Substituting the definition \eqref{eq1p11} in \eqref{eq1p13}, we find
\begin{equation}
  \dot{\mathbb{M}}(t)=\mathbb{R}(t)\mathbb{M}(t).
  \label{eq1p14}
\end{equation}
\noindent with initial conditions
\begin{equation}
  \mathbb{M}(0)=\mathds{1}.
  \label{eq1p15}
\end{equation}

However, note that the matrix $\mathbb{R}(t)$ is calculated on the classical trajectory, which in its turn is subject to boundary
conditions. Also notice that the differential equations \eqref{eq1p14} couple the blocks of the tangent matrix exclusively in
pairs. Therefore, we need to consider only the equations of motion for $M_{12}(t)$ and $M_{22} (t)$, with initial conditions
$M_{12}(0)=0$ and $M_{22}(t)=\mathds{1}$.

\subsection{Initial value representation}
\label{ssc:inivalrep}

The classical trajectory is the fundamental quantity for calculating all elements of the semiclassical propagator. However,
finding the classical solution represents a boundary condition problem, whose analytical or numerical resolution generally
exhibits greater technical difficulties or higher computational cost than a similar problem subject to initial conditions.
Therefore, the development of semiclassical propagation methods based on initial conditions, known as \textit{initial value
representations}, is highly desirable. In this section we develop such a method for \eqref{eq1p5}.

First, we use the resolution of the identity \eqref{eq1p3} to reconstruct a specific propagator from an integral over the entire
set of propagators with the same initial coherent state:
\begin{equation}
  \begin{aligned}
  K(w\cg_{f},w_{i};\tau)&=\bra w_{f}|e^{-iH\tau}|w_{i}\ket \\
  &=\intl_{\b{w}(\tau)\,\in\,\mathds{C}^{n-1}} d\mu(\b{w}\cg(\tau),\b{w}(\tau))\;
  \bra w_{f}|\b{w}\cg(\tau)\ket \bra \b{w}\cg(\tau)|e^{-iH\tau}|w_{i}\ket\\
  &=\intl_{\b{w}(\tau)\,\in\,\mathds{C}^{n-1}} d\mu(\b{w}\cg(\tau),\b{w}(\tau))\;
  \bra w_{f}|\b{w}\cg(\tau)\ket K(\b{w}(\tau),w_{i};\tau)
  \end{aligned}
  \label{eq1p16}
\end{equation}

Next we consider $\b{w}(\tau)$ as a function of the initial values of its corresponding trajectory:
\begin{equation}
  \b{w}(\tau)=\b{w}(\b{w}_{i},w_{i};\tau);
  \label{eq1p17}
\end{equation}
\noindent where $\b{w}_{i}=\b{w}(0)$. Thus, the integrand in the last line of \eqref{eq1p16} also becomes a function of
$\b{w}_{i}$ implicitly in $\b{w}(\tau)$. The change of integration variables introduces the following Jacobian determinant:
\begin{equation}
  \prodl_{j=1}^{n-1}d^{2}\b{w}_{j}(\tau)=
  \left|\det\left[\frac{\del\b{w}(\tau)}{\del\b{w}(0)}\right]\right|^{2}
  \prodl_{j=1}^{n-1}d^{2}\b{w}_{j}(0)
  =\left|\det M_{22}(\tau)\right|^{2}
  \prodl_{j=1}^{n-1}d^{2}\b{w}_{j}(0).
  \label{eq1p18}
\end{equation}

We should note that the mapping between $\b{w}_{i}$ and $\b{w}(\tau)$ is not injective, due to the existence of focal points.
However, the determinant of $M_{22}$ is zero at these problematic values of $\b{w}(\tau)$, so that their contribution to the
integral is null\footnote{In fact, as we shall see below, the focal points correspond to zeros of the whole integrand in the
initial value representation.}.

Finally, considering the semiclassical approximation for the propagators in the integrand and substituting the expression
\eqref{eq1p18} in \eqref{eq1p16}, we obtain the first form for the semiclassical propagator in the initial value representation:
\begin{equation}
  K_{sc}^{ivr}(w\cg_{f},w_{i};\tau)=\intl_{\b{w}_{i}\,\in\,
  \mathds{C}^{n-1}}d^{2}\b{w}_{i}\,
  \frac{\sigma(n)\,\dim(\mathbb{B}_{N}^{n})
  \left|\det M_{22}(\tau)\right|^{2}}{(1+\b{w}\cg(\tau) \b{w}(\tau))^{n}}
  \bra w_{f}|\b{w}\cg(\tau)\ket K_{sc}(\b{w}(\tau),w_{i};\tau);
  \label{eq1p19}
\end{equation}

\noindent where $d^{2}\b{w}_{i}=\prodl_{j=1}^{n-1}d^{2}\b{w}_{j}(0)$. Notice that the integrand of \eqref{eq1p19} is now
proportional to $\left|\det M_{22}(\tau)\right|^{\frac{3}{2}}$, instead of the inconvenient factor $\left|\det
M_{22}(\tau)\right|^{-\frac{1}{2}}$ in equation \eqref{eq1p5}. Thus we avoid the potential divergences of the semiclassical
propagator corresponding to focal points in the variables $\b{w}$. Also note that all quantities in the integrand of
\eqref{eq1p19} are calculated on the trajectory with initial conditions $w(0)=w_{i}$ and $\b{w}(0)=\b{w}_{i}$. Therefore, by
calculating the semiclassical propagator $K_{sc}(\b{w}(\b{w}_{i},w_{i};\tau),w_{i},\tau)$ for a grid of initial conditions with
$w_{i}$ fixed, we obtain the semiclassical propagator $K_{sc}^{ivr}(w\cg_{f},w_{i};\tau)$, at the desired arrival point, after an
integration in $\b{w}_{i}$.

However, our scheme to recast the propagator in terms of initial conditions seems to have some disadvantages in relation to the
original boundary condition problem. At first glance, we replaced the calculation of a single propagator by an infinite number of
propagators, which are calculated for all possible values of $\b{w}_{i}$. Even though the latter are subjected to initial
conditions, the large number of propagators in the integration can make this method impracticable. But experience tells us that
the trajectories with major contribution to the integral \eqref{eq1p19} are associated with values of $\b{w}_{i}$ close to
$w_{i}\cg$. Therefore, the integral \eqref{eq1p19} is usually calculated for a small grid around $w_{i}\cg$, considerably
reducing the number of classical trajectories required in a practical application.

The second problem in the expression \eqref{eq1p19} is the need to carry out a new integration for each choice of the final
coherent state, parametrized by $w_{f}$. However, all dependence on $w_{f}$ in the integrand of \eqref{eq1p19} comes from the
factor $\bra w_{f}|\b{w}\cg(\tau)\ket$. Hence, using the identity \eqref{eq1p2}, we can perform a multinomial expansion in the
numerator of the coherent state overlap, thus extracting $w_{f}$ from the integration sign:
\begin{equation}
  K_{sc}^{ivr}(w\cg_{f},w_{i};\tau)=\suml_{m_{1}+\ldots+m_{n}=N}
  \frac{N!}{m_{1}!\ldots m_{n}!}
  \frac{1}{(1+w\cg_{f}w_{f})^{\frac{N}{2}}}
  \left[\prodl_{j=1}^{n-1}(w\cg_{f,j})^{m_{j}}\right]
  \mathcal{I}_{m_{1},\ldots,m_{n}}
  \label{eq1p20}
\end{equation}

Hence, in order to calculate the semiclassical propagator for an arbitrary final coherent state, we need to perform only
$\dim(\mathbb{B}_{N}^{n})$ integrations whose values are independent of $w_{f}$:
\begin{equation}
  \begin{aligned}
  \mathcal{I}_{m_{1},\ldots,m_{n}}(w_{i};\tau)&=\int d^{2}\b{w}_{i}\,
  \frac{\sigma(n)\,\dim(\mathbb{B}_{N}^{n})\left|\det M_{22}(\tau)\right|^{2}
  \prodl_{j=1}^{n-1}\left[\b{w}\cg_{j}(\tau)\right]^{m_{j}}}
  {(1+\b{w}\cg(\tau)\b{w}(\tau))^{\frac{N}{2}+n}}
  K_{sc}(\b{w}(\tau),w_{i};\tau)\\
  &=\left(\frac{N!}{m_{1}!m_{2}!\ldots m_{n}!}\right)^{-\frac{1}{2}}\left.\bra
  m_{1},m_{2},\ldots,m_{n}|e^{-iH\tau}|w_{i}\ket\right|_{sc}.
  \end{aligned}
  \label{eq1p21}
\end{equation}

The second equality shows that the integrals $\mathcal{I}_{m_{1},\ldots,m_{n}}$ can be rewritten as semiclassical propagators
between the initial coherent state and a number state, except by a combinatorial factor.

\subsection{Heuristic filters}
\label{sc:heurstfilt}

It is well known that some trajectories in the doubled phase space give unphysical contributions to the semiclassical
propagator\cite{Adachi89, Rubin95, Shudo95, Shudo96, Aguiar05, Aguiar10}. Therefore, given a grid of initial conditions
$\b{w}_{i}$, only part of the resulting classical trajectories participate in the calculation of the integrals \eqref{eq1p21}.
The appropriate contributions can be collected using the heuristic filter defined by:
\begin{equation}
  \frac{d}{dt}\Ln\left(\left|K_{sc}(\b{w}(t),w_{i};t)\right|^{2}\right)<\lambda.
  \label{eq1p22}
\end{equation}

The classical trajectories that violate this condition at time $t$ are discarded from the integration for $\tau>t$. Note that the
only free parameter in the initial value representation is $\lambda$, whose positive value should be adjusted in order to
optimize the semiclassical propagation.

The idea behind this filter is the following: if we write the semiclassical propagator as $K_{sc}=e^{\alpha+i\beta}$, with
$\alpha,\beta\in\mathds{R}$, then the inequality \eqref{eq1p22} can be recast in the form $\frac{d\alpha}{dt}<\frac{\lambda}{2}$.
Therefore, the discarded trajectories are those that lead to an abrupt positive change in the real part of
$\Ln\left(K_{sc}\right)$, thus causing the divergence of the absolute value of the propagator. As seen in the equation
\eqref{eq1p5}, the time variations in $\alpha$ are directly determined by the imaginary part of the corrected action $(S+I)$.
However, unlike previously published methods\cite{Rubin95, Aguiar10}, the proposed heuristic filter also takes into account the
factor that contains the determinant of the tangent matrix. Clearly, the modulus of this factor also affects the value of
$\alpha$, either counteracting abrupt negative changes in $\Im(S+I)$ or contributing to the divergence of the semiclassical
propagator. The inclusion of this aspect in the heuristic filter is an important element in the present work, which greatly
improved the results in section \ref{sc:appldisc}.

\subsection{$Q$ representation with $\mathrm{SU}(n)$ coherent states}
\label{ssc:qrepsun}

Using the expressions \eqref{eq1p20} and \eqref{eq1p21}, we can easily calculate the semiclassical propagator at any point $w$ of
the \textit{classical phase space}\footnote{Note that, for simplicity of notation, we omit the subindex `$f$' for the final
condition of the semiclassical propagator. In this way we also emphasize the role of the variables $w$ as coordinates of a
classical phase space in which we can represent the quantum states and operators.}, for fixed initial condition $w_{i}$ and
period of propagation $\tau$. Thus, we obtain a complete description of the system state, known as the Husimi or $Q$
representation\cite{Scully97}. In general, the function $Q(w\cg,w)$ associated with an arbitrary state $|\psi\ket$ is defined as:
\begin{equation}
  \begin{aligned}
  Q(w\cg,w)&=\bra w|\rho|w\ket\\
  &=|K(w\cg,w_{i};\tau)|^2;
  \end{aligned}
  \label{eq1p23}
\end{equation}

\noindent where $\rho=|\psi\ket\bra\psi|$ is the density operator for a pure state and $|w\ket$ is given by equation
\eqref{eq1p1}. In the second line of \eqref{eq1p23} we assume that $|\psi\ket=e^{-iH\tau}|w_{i}\ket$. Therefore, using the
$\mathrm{SU}(n)$ semiclassical propagator, we can directly construct the semiclassical representation
$Q_{sc}(w\cg,w)=|K_{sc}(w\cg,w_{i},\tau)|^2$.

With the aid of the expression \eqref{eq1p3} and assuming $\bra\psi|\psi\ket=1$ we find that:
\begin{equation}
  \intl_{w\,\in\,\mathds{C}^{n-1}} d\mu(w\cg,w)\;Q(w\cg,w)=1.
  \label{eq1p24}
\end{equation}

Unlike the exact definition \eqref{eq1p4}, the semiclassical propagators \eqref{eq1p5} and \eqref{eq1p20} do not preserve the
norm of the state during its evolution\cite{Aguiar10}. Therefore, for a proper comparison with the quantum results at time
$\tau$, we need to normalize $Q_{sc}$ according to the relation \eqref{eq1p24}. The normalization of the quantum and
semiclassical representations is implied in the remainder of the paper.

In terms of the exact $Q$ function or of its semiclassical version $Q_{sc}$, we can readily obtain the mean of an arbitrary
observable $O$:
\begin{equation}
  \bra O\ket=\int d\mu(w\cg,w) \mcl{O}_{\mathrm{a}}(w\cg,w)Q(w\cg,w).
  \label{eq1p25}
\end{equation}

The function $\mcl{O}_{\mathrm{a}}$, which corresponds to the antinormally ordered symbol of the operator $O$, is defined by
\begin{equation}
  O=\int d\mu(w\cg,w) \mcl{O}_{\mathrm{a}}(w\cg,w)|w\ket\bra w|.
  \label{eq1p26}
\end{equation}

\section{Application and discussion of the semiclassical propagator}
\label{sc:appldisc}

\subsection{Bose-Einstein condensate in a triple-well trapping potential}
\label{ssc:bosetripwell}

In order to illustrate the method described in previous sections, we discuss here its application to $\mathrm{SU}(2)$ and
$\mathrm{SU}(3)$ coherent states, considering a simplified model for the dynamics of a Bose-Einstein condensate in a triple-well
potential\cite{Viscondi11a}. Assuming that the three wells of the trap are identical and equivalently coupled, the Hamiltonian of
the model in a three-mode approximation is given by:
\begin{equation}
  H=\Omega\suml_{j\neq k}a\dg_{j}a_{k}+\frac{\chi}{(N-1)}\suml_{j=1}^{3}(a\dg_{j})^{2}a_{j}^{2};
  \label{eq2p1}
\end{equation}

\noindent where $a_{j}$ ($a\dg_{j}$) is the bosonic annihilation (creation) operator related to the single-particle state
$|u_{j}\ket $, which represents the ground state of a harmonic oscillator centered on $j$-th minimum of the trapping potential,
for $j=1,2,3$. The parameters $\Omega$ and $\chi$ correspond to the rates of tunneling and collision of trapped bosons,
respectively.

Note that $H$ preserves the total number of particles, so that we can restrict our analysis to invariant subspaces with fixed
$N$, denoted by $\mathds{B}^{3}_{N}$. Therefore, the $\mathrm{SU}(3)$ coherent states, defined in equation \eqref{eq1p1} with
$n=3$, are appropriate to study the model. Substituting \eqref{eq2p1} in \eqref{eq1p7}, we obtain the effective classical
Hamiltonian:
\begin{equation}
  \begin{aligned}
  \frac{\mcl{H}}{N}=&
  \Omega\frac{\b{w}_{1}w_{2}+\b{w}_{2}w_{1}+\b{w}_{1}+w_{1}+\b{w}_{2}+w_{2}}
  {1+\b{w}_{1}w_{1}+\b{w}_{2}w_{2}}\\
  &+\chi\frac{\b{w}_{1}^{2}w_{1}^{2}+\b{w}_{2}^{2}w_{2}^{2}+1}
  {(1+\b{w}_{1}w_{1}+\b{w}_{2}w_{2})^{2}}.
  \end{aligned}
  \label{eq2p2}
\end{equation}

Then, employing the general formula \eqref{eq1p6}, we find the classical equations of motion for the condensate:
\begin{equation}
  \begin{aligned}
  i\dot{w}_{j}&=\Omega(1+w_{1}+w_{2})(1-w_{j})
  +2\chi\frac{w_{j}(\b{w}_{j}w_{j}-1)}{1+\b{w}_{1}w_{1}+\b{w}_{2}w_{2}}\\
  -i\dot{\b{w}}_{j}&=\Omega(1+\b{w}_{1}+\b{w}_{2})(1-\b{w}_{j})
  +2\chi\frac{\b{w}_{j}(\b{w}_{j}w_{j}-1)}{1+\b{w}_{1}w_{1}+\b{w}_{2}w_{2}}
  \end{aligned}
  \label{eq2p3}
\end{equation}
\noindent for $j=1,2$. Using \eqref{eq2p2} and \eqref{eq2p3}, we can easily obtain the other dynamical quantities relevant to the
calculation of the semiclassical propagator, such as the Lagrangian $L$ and the matrix $\mathbb{R}$. According to the equations
\eqref{eq2p3}, the dynamics of the condensate exhibits three \textit{classical invariant subspaces}, described by the following
conditions:
\begin{subequations}
  \begin{align}
  &w_{1}=w_{2},\;\b{w}_{1}=\b{w}_{2}\label{eq2p4a}\\
  &w_{1}=1,\;\b{w}_{1}=1\label{eq2p4b}\\
  &w_{2}=1,\;\b{w}_{2}=1\label{eq2p4c}
  \end{align}
  \label{eq2p4}
\end{subequations}
For simplicity, we limit our discussion to the case \eqref{eq2p4a}, since the three invariant subspaces are dynamically
equivalent\cite{Viscondi11a}. Now, we show that the effective quantum dynamics of the condensate under the constraints
\eqref{eq2p4} can be approximated by $\mathrm{SU}(2)$ semiclassical propagators. For this purpose, we rewrite the coherent state
\eqref{eq1p1} in terms of bosonic creation operators:
\begin{equation}
  |w\ket=\frac{1}{\sqrt{N!}}\left[\begin{array}{c}\displaystyle{
  \frac{\suml_{j=1}^{n-1}w_{j}a\dg_{j}+a\dg_{n}}{(1+w\cg w)^{\frac{1}{2}}}}
  \end{array}\right]^{N}|0\ket.
  \label{eq2p5}
\end{equation}

Then, we apply the condition \eqref{eq2p4a} to the equation \eqref{eq2p5} for $n=3$:
\begin{equation}
  \begin{aligned}
  |w_{1},w_{2}=w_{1}\ket_{_{SU(3)}}&=\frac{1}{\sqrt{N!}}
  \left[\frac{w_{1}(a\dg_{1}+a\dg_{2})+a\dg_{3}}{(1+2w\cg_{1}w_{1})^{\frac{1}{2}}}\right]^{N}|0\ket\\
  &=\frac{1}{\sqrt{N!}}\left[\frac{\sqrt{2}w_{1}b\dg_{1}+b\dg_{2}}{(1+2w\cg_{1}w_{1})^{\frac{1}{2}}}\right]^{N}|0\ket\\
  &=|\sqrt{2}w_{1}\ket_{_{SU(2)}};
  \end{aligned}
  \label{eq2p6}
\end{equation}
\noindent where we performed a change of basis in the single-particle Hilbert space, corresponding to the following unitary
transformation of the bosonic creation operators\cite{Negele98}:
\begin{equation}
  \left\{
  \begin{aligned}
  b\dg_{1}&=\frac{1}{\sqrt{2}}\left(a\dg_{1}+a\dg_{2}\right)\\
  b\dg_{2}&=a\dg_{3}\\
  b\dg_{3}&=\frac{1}{\sqrt{2}}\left(a\dg_{1}-a\dg_{2}\right)
  \end{aligned}
  \right.
  \label{eq2p7}
\end{equation}

According to the equation \eqref{eq2p6}, when restricted to a invariant subspace under the classical dynamics, the
$\mathrm{SU}(3)$ coherent states are reduced to the $\mathrm{SU}(2)$ coherent states with parameter $\sqrt{2}w_{1}$.

Also notice that the state presented in \eqref{eq2p6} has zero occupation number in the mode associated with the operator
$b\dg_{3}$. Therefore, the constraint \eqref{eq2p4a} is \textit{classically} equivalent to the equation $\bra
b\dg_{3}b_{3}\ket=0$. However, by applying the transformation \eqref{eq2p7} to the Hamiltonian \eqref{eq2p1}, we can easily see
that the mean occupation $\bra b\dg_{3}b_{3}\ket$ does not remain zero under the \textit{quantum} evolution of the condensate,
considering any state initially unoccupied in this mode. Consequently, the subspaces \eqref{eq2p4} do not have quantum
counterparts with identical characteristics. However, we can still use the $\mathrm{SU}(2)$ coherent states to approximate the
\textit{semiclassical} dynamics under these restrictions. This approximation should provide accurate results when a similar
evolution in the unrestricted space displays irrelevant values of $\bra b\dg_{3}b_{3}\ket$.

\subsection{Classical approximation}
\label{ssc:classapprox}

In order to establish a criterion for comparison between the semiclassical and quantum results, we now introduce a third approach
to the bosonic dynamics, which we call \textit{classical approximation}.

We designate as \textit{principal trajectory}, indicated by the subindex `$p$', the solution of the classical equations of motion
\eqref{eq1p6} subject to initial conditions $w_{p}(0)=w_{i}$ and $\b{w}_{p}(0)=w\cg_{i}$. In this case the two vector equations
in \eqref{eq1p6} become redundant, since the solution is such that $w_{p}(t)=\b{w}\cg_{p}(t)$\footnote{Notice that the action $S$
and the correction term $I$ are real valued when calculated on the principal trajectory. This property makes the removal of the
principal trajectory by the heuristic filter a very unlikely event, as can be inferred from the discussion below the inequality
\eqref{eq1p22}.}.

The classical approximation to the mean of an arbitrary observable $O$ at the time $\tau$ is defined as follows:
\begin{equation}
  \begin{aligned}
  \bra O\ket_{c}(\tau)&=\bra w_{p}(\tau)|O|w_{p}(\tau)\ket;
  \end{aligned}
  \label{eq2p11}
\end{equation}
\noindent where $|w_{p}(\tau)\ket$ indicates the coherent state parametrized by the principal trajectory $w_{p}(\tau)$. The
classical approximation consists simply in calculating the function $\mcl{O}_{\mathrm{n}}(w\cg,w)=\bra w|O|w\ket$, which
represents the normally ordered symbol of the operator $O$, on the principal trajectory.

Assuming an initial state $|w_{i}\ket_{_{\mathrm{SU}(n)}}$, the classical approximation of $\bra O\ket(t)$ is exact in only two
specific situations when compared with the corresponding quantum results: (\textrm{i}) for $H\in\mathrm{su}(n)$, because in this
case $|w_{p}(\tau)\ket$ differs from the correct solution of the Schrödinger equation by no more than a global
phase\cite{Zhang90}; (\textrm{ii}) in the macroscopic limit, given by $N\rightarrow\infty$\cite{Yaffe82}.

Clearly, the semiclassical approximation is more accurate than the classical approach \eqref{eq2p11}, since it adds quantum
corrections to the classical results. Therefore, the semiclassical propagator \eqref{eq1p5} is also exact for any linear
Hamiltonian in the generators of $\mathrm{SU}(n)$ ($H\in\mathrm{su}(n)$) as well as in the macroscopic limit
($N\rightarrow\infty$).

Under the restriction $H\in\mathrm{su}(n)$, every initial condition $\b{w}_{i}$ must provide a trajectory with appropriate
contribution to the integral \eqref{eq1p21}. Accordingly, the heuristic filter \eqref{eq1p22} must allow the contribution of all
trajectories at all instants of time, which it does, because $\left|K_{sc}(\b{w}(t),w_{i};t)\right|^{2}$ is constant with respect
to $t$ for linear Hamiltonians.

It follows that the classical and semiclassical approximations to the Hamiltonian \eqref{eq2p1} are exact for $\chi=0$, since in
this regime $H$ is linear in the generators of $\mathrm{SU}(3)$ (bilinear in the creation and annihilation operators). Therefore,
the bosonic collisions introduce nonlinear terms to the condensate dynamics, whose classical and semiclassical descriptions are
not complete for a finite number of particles. Consequently, we expect the application of the semiclassical propagator
\eqref{eq1p20} to be better behaved for weak nonlinearities (small values of $\chi$) and large numbers of bosons.

\subsection{Semiclassical approximation with $\mathrm{SU}(2)$ coherent states}
\label{ssc:appsu2}

A relevant observable in the condensate dynamics is the population imbalance operator $S_{z}$, which describes the difference in
occupation between the two effectively occupied modes in the classical invariant subspace \eqref{eq2p4a}:
\begin{equation}
  S_{z}=\frac{b\dg_{1}b_{1}-b\dg_{2}b_{2}}{2}
  \label{eq2p12}
\end{equation}

\begin{figure}[htbp]
  \centering
  \includegraphics[width=0.5\linewidth]{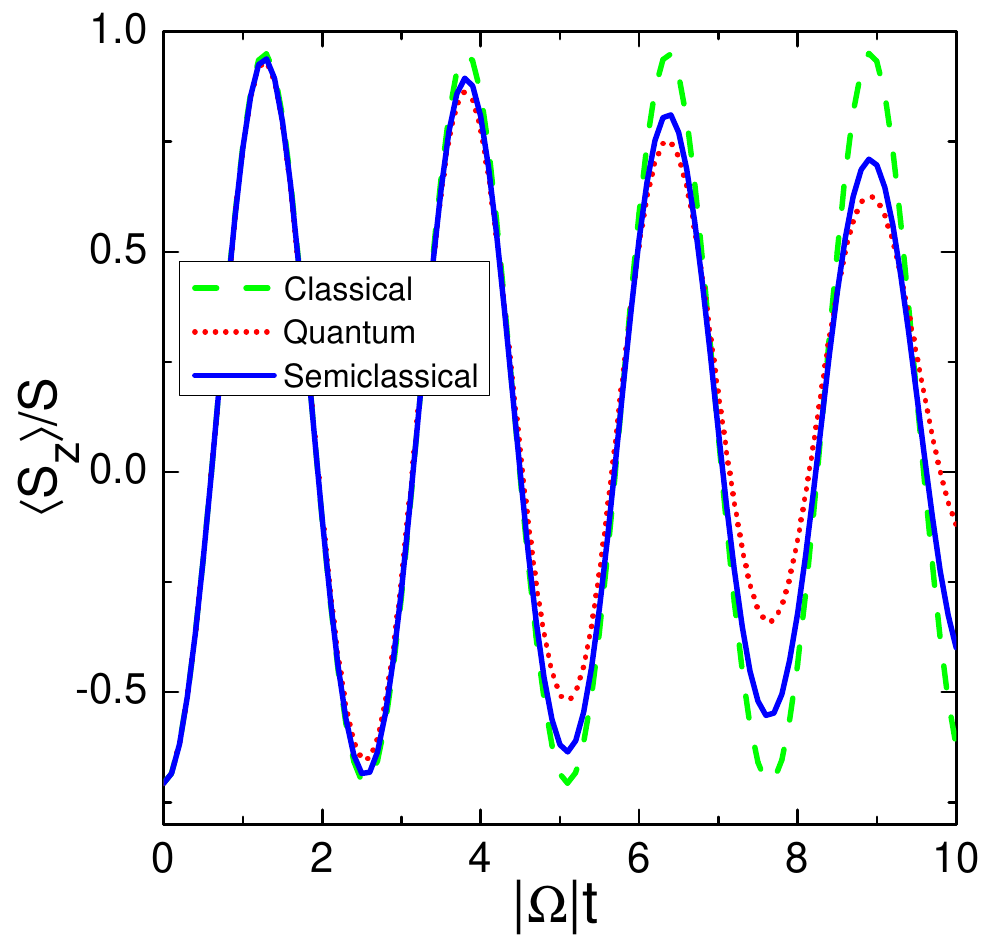}
  \caption{Classical (dashed green), quantum (dotted red) and semiclassical (solid blue) evolution of the normalized
  mean of the population imbalance operator $S_{z}$ for the initial state $|\sqrt{2}w_{1}\ket_{_{\mathrm{SU}(2)}}=
  |\tan\frac{\pi}{8}\ket_{_{\mathrm{SU}(2)}}$. The parameters of the Hamiltonian assume the values $N=2S=30$,
  $\Omega=-1$ and $\chi=-1$. The $\mathrm{SU}(2)$ semiclassical propagation was performed with a grid of
  $479$ initial conditions and limiting filter $\lambda=10$.}
  \label{fig1}
\end{figure}

Figure \ref{fig1} compares the semiclassical, quantum and classical evolution of $\bra S_{z}\ket/S$ for $N=30$, $\Omega=-1$ and
$\chi=-1$, considering as initial state $|\sqrt{2}w_{1}\ket_{_{\mathrm{SU}(2)}}=|\tan\frac{\pi}{8}\ket_{_{\mathrm{ SU}(2)}}$. The
mean of $S_{z}$ is normalized by the quantity $S=\frac{N}{2}$ so that $-1\leq\bra S_{z}\ket/S\leq1$. For the semiclassical
approximation we used the $\mathrm{SU}(2)$ propagator with $479$ initial conditions $\b{w}_{i}$ and limiting value $\lambda=10$
for the heuristic filter.

Notice that the oscillations of the classical mean display constant amplitude, unlike the semiclassical and quantum results.
Although restricted to the $\mathrm{SU}(2)$ propagator, the semiclassical method shows quantitative agreement with the exact
quantum calculations, being fairly superior to the classical approximation, even for a relatively small number of particles. In
general, the classical and semiclassical approximations are accurate for sufficiently short times, but the quality of the
semiclassical evolution is obviously higher for longer periods of propagation, when the nonlinear terms of the quantum
Hamiltonian become important.

\begin{figure}[htbp]
  \centering
  \includegraphics[width=0.5\linewidth]{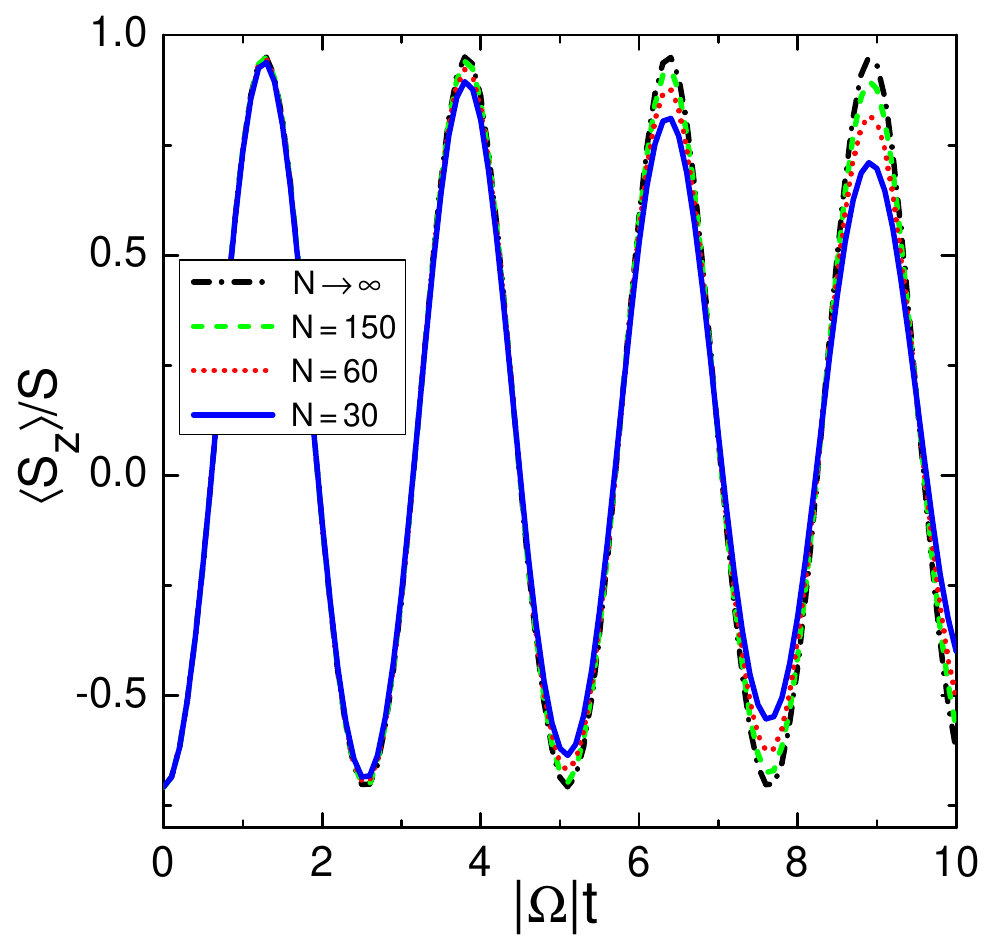}
  \caption{Semiclassical evolution of $\bra S_{z}\ket/S$ for $N=30$ (solid blue), $N=60$ (dotted red) and $N=150$
  (dashed green). The dash-dotted black curve represents the classical approximation, which is equivalent to the
  macroscopic limit $N\rightarrow\infty$. In all results we consider $\Omega=-1$, $\chi=-1$ and initial state
  $|\sqrt{2}w_{1}\ket_{_{\mathrm{SU}(2)}} = |\tan\frac{\pi}{8}\ket_{_{\mathrm {SU}(2)}}$.}
  \label{fig2}
\end{figure}

Figure \ref{fig2} shows the behavior of the semiclassical evolution of $\bra S_{z}\ket/S$ with the variation of the total number
of particles, for $\Omega=-1$, $\chi=-1$ and initial state $|\sqrt{2}w_{1}\ket_{_{\mathrm{SU}(2)}} =
|\tan\frac{\pi}{8}\ket_{_{\mathrm{SU}(2)}}$. The results correspond to the $\mathrm{SU}(2)$ semiclassical propagator for $30$,
$60$ and $150$ particles, with $\lambda=10$ and about $500$ initial conditions in each case.

Note that the equations of motion \eqref{eq2p3} and their solutions, including the principal trajectory $w_{p}(t)$, are
independent of the total number of particles. Therefore, it is easy to show that, for a linear operator in the generators of
$\mathrm{SU}(3)$, the \textit{classical mean per particle} is also independent of $N$. Therefore, quantities like $\bra
S_{z}\ket_{c}/S$ represent the macroscopic limit of their quantum and semiclassical counterparts, since the classical
approximation \eqref{eq2p11} is exact for $N\rightarrow\infty$.

In accordance with the previous discussion, we included the classical approximation in figure \ref{fig2} as the macroscopic limit
for the dynamics of the semiclassical means. Note that the semiclassical results quickly converge to the classical curve with
increasing $N$. Consequently, we expect the classical approximation to show high accuracy for a few hundred condensate bosons,
which represents a scenario compatible with usual experiments. However, the semiclassical propagators must provide superior
results for the mesoscopic dynamics when subjected to longer periods of propagation or more intense nonlinear effects.

\begin{figure}[htbp]
  \centering
  \includegraphics[width=0.5\linewidth]{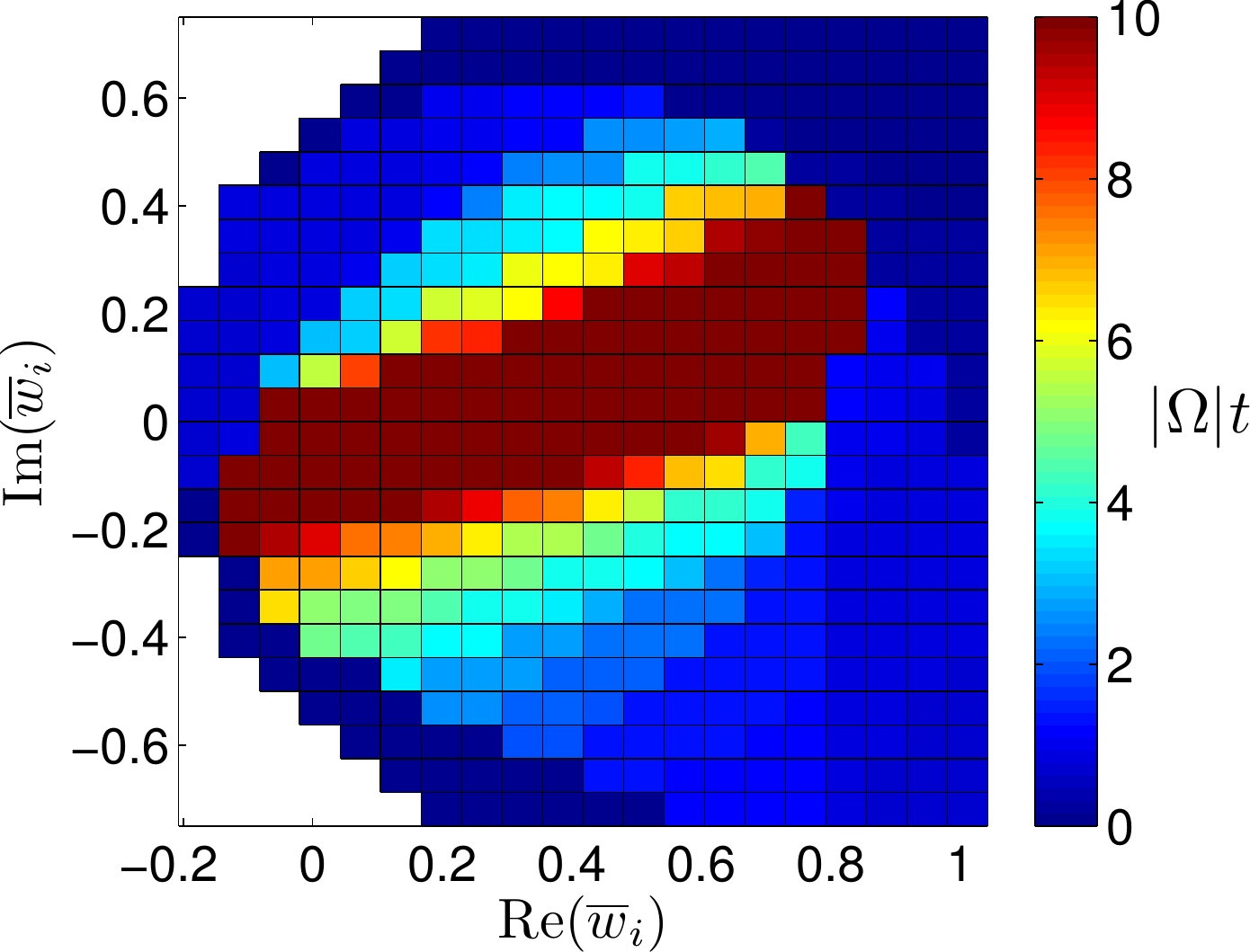}
  \caption{Diagram of contributing trajectories for the $\mathrm{SU}(2)$ semiclassical propagator
  with $N=30$, $\Omega=-1$, $\chi=-1$ and initial state $|\sqrt{2}w_{1}\ket_{_{\mathrm{SU}(2)}} =
  |\tan\frac{\pi}{8}\ket_{_{\mathrm{SU}(2)}}$. Each initial condition $\b{w}_{i}$ is
  represented by a square whose color indicates the period of contribution of the corresponding
  trajectory, according to the heuristic filter \eqref{eq1p22} for $\lambda=10$.}
  \label{fig3}
\end{figure}

Figure \ref{fig3} shows the diagram of contributing trajectories for the $\mathrm{SU}(2)$ semiclassical propagator with $N=30$,
$\Omega=-1$, $\chi=-1$ and initial state $|\sqrt{2}w_{1}\ket_{_{\mathrm{SU}(2)}} = |\tan\frac{\pi}{8}\ket_{_{\mathrm{SU}(2)}}$.
This diagram corresponds to the semiclassical approximation shown in figure \ref{fig1} and reproduced in figure \ref{fig2}. Each
square in figure \ref{fig3} represents an initial condition $\b{w}_{i}$ used in the numerical calculation of the integrals
\eqref{eq1p21}. The color code indicates the time of contribution of the resulting classical trajectories, determined by the
heuristic filter \eqref{eq1p22} with $\lambda=10$.

Notice that the trajectories with the most significant contributions have initial conditions $\b{w}_{i}$ centered around
$w_{i}\cg=\frac{1}{\sqrt{2}}\tan\frac{\pi}{8}\approx0.29$. This initial value defines the principal trajectory, whose
contribution is among the most important in the reconstruction of the semiclassical propagator. Note also that $w\cg_{i}$ is the
value that maximizes the $Q$ representation for the state $|\sqrt{2}w\cg_{i}\ket_{_{\mathrm {SU}(2)}}$. Therefore, this coherent
state is located in the same region of phase space responsible for the most relevant contributions to the initial value
representation.

\begin{figure}[htbp]
  \centering
  \includegraphics[width=0.5\linewidth]{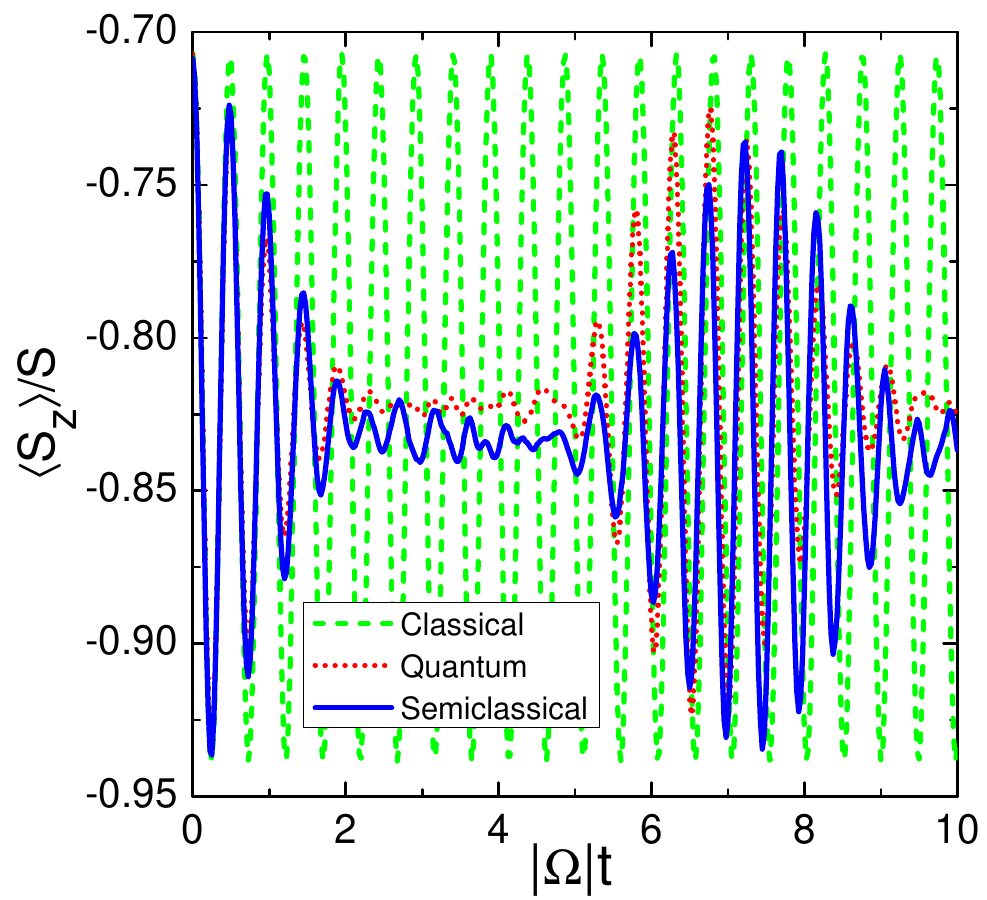}
  \caption{Classical (dashed green), quantum (dotted red) and semiclassical (solid blue) evolution of
  $\bra S_{z}\ket/S$ for $N=30$, $\Omega=-1$, $\chi=-8$ and coherent initial state
  $|\sqrt{2}w_{1}\ket_{_{\mathrm{SU}(2)}} = |\tan\frac{\pi}{8}\ket_{_{\mathrm{SU}(2)}}$. The semiclassical
  approximation was performed with the $\mathrm{SU}(2)$ propagator, considering a grid of $3781$
  initial conditions and $\lambda=18$.}
  \label{fig4}
\end{figure}

In general, the equations of motion resulting from the Hamiltonian \eqref{eq2p1} show significant changes in behavior for
different magnitudes of the ratio $\chi/\Omega$\cite{Viscondi10, Viscondi11a}, which represents the relative intensity between
the quadratic and linear terms of $H$. The previous examples of application of the semiclassical propagator are restricted to
small absolute values of $\chi/\Omega$, since the linear terms are clearly dominant in the dynamics of the condensate. Figure
\ref{fig4} displays the semiclassical, quantum and classical dynamics of $\bra S_{z}\ket/S$ in a strongly nonlinear regime, for
$N=30$, $\Omega=-1$, $\chi=-8$ and initial state $|\sqrt{2}w_{1}\ket_{_{\mathrm{SU}(2)}} =
|\tan\frac{\pi}{8}\ket_{_{\mathrm{SU}(2)}}$. In the semiclassical approximation, we employed the $\mathrm{SU}(2)$ propagator for
a grid of $3781$ initial conditions and limiting value $\lambda=18$.

Again we see that the amplitude of the classical mean remains constant during the whole evolution of the system. Conversely, the
semiclassical and quantum results exhibit an almost complete `collapse' of the oscillations, followed by a partial `revival' of
the amplitude value in relation to the classical approximation. Therefore, this example refers to a strongly nonlinear and
exclusively quantum behavior, described with excellent accuracy by the semiclassical propagator. However, note that the number of
trajectories required for a proper semiclassical approximation is considerably larger than in the predominantly linear dynamics
shown in figure \ref{fig1}. As expected, the semiclassical propagator loses computational efficiency in nonlinear regimes.

\begin{figure*}[htbp]
  \centering
  \includegraphics[width=\linewidth]{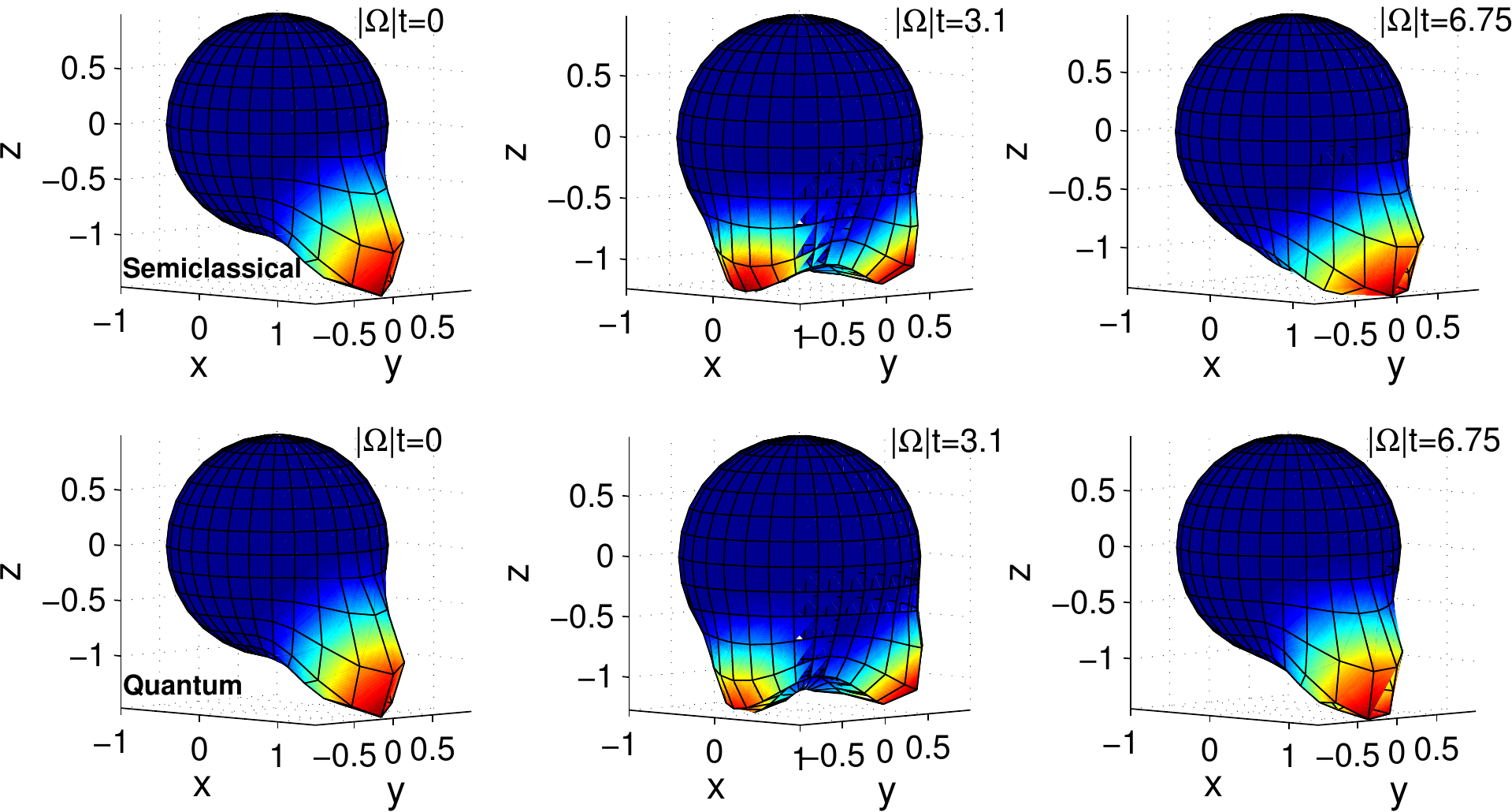}
  \caption{At the top (bottom) we show the $Q$ representation on the unit sphere related to the semiclassical (quantum)
  evolution of condensate at three different times, for $N=30$, $\Omega=-1$, $\chi=-8$ and initial state
  $|\sqrt{2}w_{1}\ket_{_{\mathrm{SU}(2)}} = |\tan\frac{\pi}{8}\ket_{_{\mathrm{SU}(2)}}$.}
  \label{fig5}
\end{figure*}

The phase space corresponding to the $\mathrm{SU}(2)$ coherent states may be identified as a spherical surface\cite{Arecchi72}.
It follows that, applying the definition \eqref{eq1p23} with the coherent states given by \eqref{eq2p6} under the transformation
of variables $\sqrt{2}w_{1} = e^{-i\phi}\tan\frac{\theta}{2}$, we obtain the $Q$ representation for $\mathrm{SU}(2)$ in terms of
angular spherical coordinates. In this way, we can represent an arbitrary quantum state on the unit sphere:
\begin{equation}
  \left\{\
  \begin{aligned}
  &x=[Q(\theta,\phi)+1]\sin\theta\cos\phi\\
  &y=[Q(\theta,\phi)+1]\sin\theta\sin\phi\\
  &z=-[Q(\theta,\phi)+1]\cos\theta
  \end{aligned}
  \right.
  \label{eq2p13}
\end{equation}
\noindent where $\phi\in[0,2\pi)$ and $\theta=[0,\pi]$. Notice that our definition for the variable $\theta$ has its origin in
the negative $z$ semi-axis.

In figure \ref{fig5} we show the comparison between the semiclassical (top) and quantum (bottom) $Q$ representations at three
different times, for $N=30$, $\Omega=-1$, $\chi=-8$ and initial state $|\sqrt{2}w_{1}\ket_{_{\mathrm{SU}(2)}} =
|\tan\frac{\pi}{8}\ket_{_{\mathrm{SU}(2)}}$. The represented states are in correspondence with the results displayed in figure
\ref{fig4}.

At $|\Omega|t=0$ we show the initial coherent state, whose representation is identical in the semiclassical and quantum
approaches. At the time $|\Omega|t=3.1$, we have the superposition of two \textit{localized} states in phase space
(`Schrödinger-cat' state), which is responsible for the oscillation collapse in $\bra S_{z}\ket/S$. At $|\Omega|t=6.75$, we see
that the $Q$ function converges again to a single location on the sphere. This behavior is associated with the revival of the
oscillations in figure \ref{fig4}.

The differences between the quantum and semiclassical representations in figure \ref{fig5} are almost imperceptible, evidencing
that the semiclassical approximation accurately describes the \textit{delocalization} and the subsequent \textit{relocalization}
of the state in the phase space.

\subsection{$\mathrm{SU}(3)$ semiclassical propagator}
\label{ssc:su3semiprop}

\begin{figure}[htbp]
  \centering
  \includegraphics[width=0.5\linewidth]{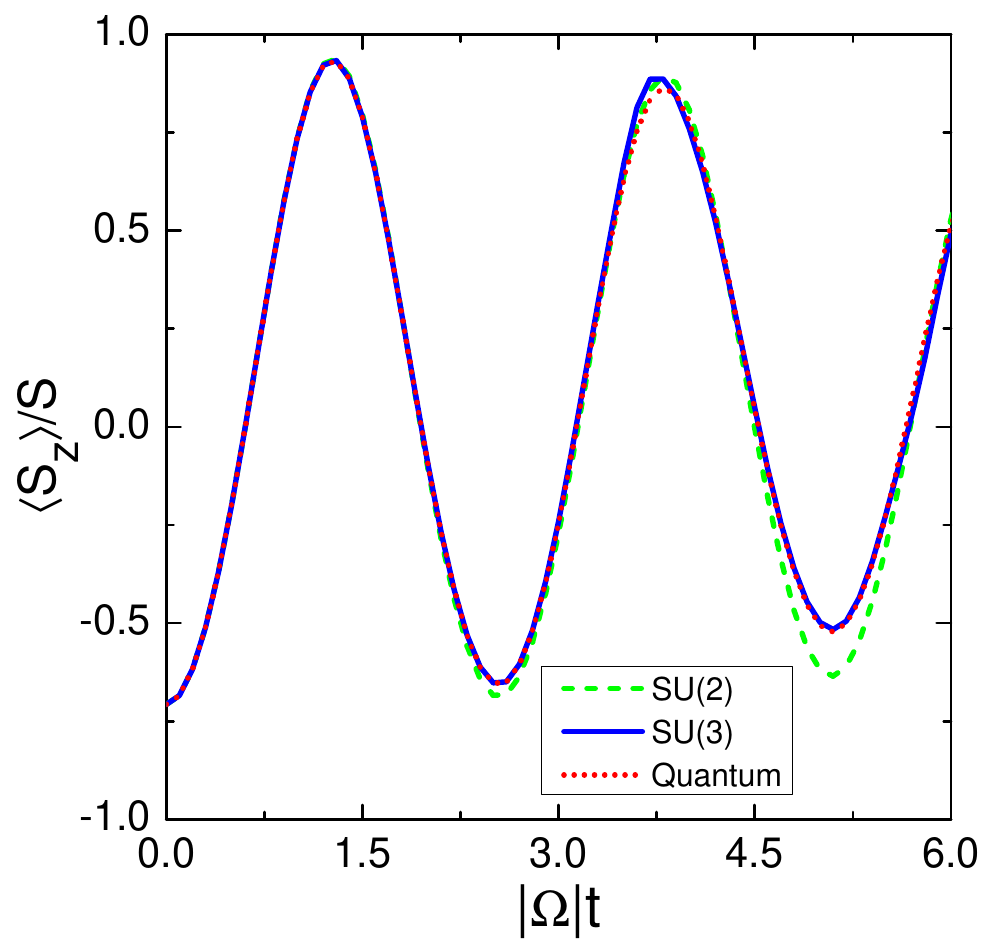}
  \caption{Time evolution of $\bra S_{z}\ket/S$ resulting from the $\mathrm{SU}(2)$ semiclassical propagator (dashed green),
  the $\mathrm{SU}(3)$ semiclassical propagator (solid blue) and the exact quantum calculation (dotted red), for $N=30$,
  $\Omega=-1$, $\chi=-1$ and initial coherent state parametrized by $w_{1}=w_{2}=\frac {1}{\sqrt{2}}\tan\frac{\pi}{8}$.
  In the $\mathrm{SU}(3)$ semiclassical approximation we used $35134$ classical trajectories, with contributing period
  determined by $\lambda=10$. The $\mathrm{SU}(2)$ semiclassical curve is the same one shown in figure \ref{fig1}.}
  \label{fig6}
\end{figure}

Although the approximations with the $\mathrm{SU}(2)$ semiclassical propagator have shown excellent accuracy, the
$\mathrm{SU}(3)$ coherent states are more appropriate to the dynamics determined by the Hamiltonian \eqref{eq2p1}. Figure
\ref{fig6} exemplifies the use of the $\mathrm{SU}(3)$ semiclassical propagator in the evolution of $\bra S_{z}\ket/S$, for
$N=30$ $\Omega=-1$, $\chi=-1$ and initial coherent state parametrized by $w_{1}=w_{2}=\frac{1}{\sqrt{2}}\tan\frac{\pi}{8}$. In
the calculation of the initial value representation we used $35134$ classical trajectories, whose contributions were determined
by the heuristic filter \eqref{eq1p22} with $\lambda=10$. In comparison with the result for the $\mathrm{SU}(3)$ propagator, we
reproduce in figure \ref{fig6} the corresponding $\mathrm{SU}(2)$ approximation and the exact quantum evolution, also shown in
figure \ref{fig1}.

As expected, the $\mathrm{SU}(3)$ semiclassical propagator is more accurate than the $\mathrm{SU}(2)$ approximation. The
difference between these results comes mainly from the occupation of the mode associated with the operator $b\dg_{3}$. During the
considered period of propagation, the normalized mean $\bra b\dg_{3}b_{3}\ket/N$ grows monotonically until it reaches a value
close to $0.04$ at $|\Omega|t=6$.

We conclude that most of the inaccuracy attributed to the $\mathrm{SU}(2)$ semiclassical propagator in figures \ref{fig1} and
\ref{fig6} is due to the classical constraint \eqref{eq2p4a}, since the $\mathrm{SU}(3)$ semiclassical approximation is almost
exact in the predominantly linear dynamical regime.

\section{Conclusion}
\label{sc:conclu}

We constructed an initial value representation for the $\mathrm{SU}(n)$ semiclassical propagator, which replaces the search for
boundary-valued trajectories by an integral over a set of initial-valued trajectories in the doubled phase space. This
formulation represents a considerable advantage in the calculation of the propagator, since the numerical or analytical
resolution of a boundary condition problem is typically much more difficult than its initial condition counterpart, particularly
in systems with many degrees of freedom. Moreover, our method allows the factorization of the arrival point $w\cg_{f}$, as given
in equation \eqref{eq1p20}, considerably reducing the number of integrations required for a complete representation of the
system.

The semiclassical approach showed excellent accuracy when compared to exact quantum results, even for a relatively small number
of particles. The efficacy of the semiclassical approximation is largely due to the effective heuristic filter, which was able to
discriminate the trajectories with appropriate contributions to the propagator. The systematic elimination of non-contributing
trajectories represents a crucial component in the implementation of an initial value representation in the doubled phase space,
because it directly determines the speed, precision and applicability of the method.

We tested our semiclassical formula for a triple-well Bose-Einstein condensate in nonlinear and predominantly linear dynamical
regimes. Although the semiclassical propagation has been very satisfactory in both situations, the number of initial conditions
required for an appropriate description of the nonlinear dynamics is significantly higher than in the almost linear case. In
general, the computational efficiency of the semiclassical propagator is only limited by the required number of contributing
classical trajectories. Clearly, this number \textit{grows} with a exponent proportional to $(n-1)$, the dimension of the
subspace $\b{w}$. However, we can assume that the required number of initial conditions $\b{w}_{i}$ \textit{decreases} with the
total number of particles, since the semiclassical results converge with increasing $N$ to the classical approximation, which is
determined by a single trajectory. Therefore, the $\mathrm{SU}(n)$ semiclassical propagator is a viable alternative in the study
of bosonic systems with many degrees of freedom and large number of particles, since the computational cost of exact quantum
methods typically grows as a polynomial in $N$ of order proportional to $n$.

Finally, we would like to point out that the formulas \eqref{eq1p20} and \eqref{eq1p21}, the main results of this paper, can be
easily extended to other classes of coherent states, such as the usual harmonic-oscillator coherent states. Thus, this work also
represents an alternative to previously published semiclassical methods.

\begin{acknowledgments}
We acknowledge the financial support from CNPq and FAPESP, under grants No. 2008/09491-9 and 2009/11032-5.
\end{acknowledgments}

\bibliography{AplPropag}

\begin{thebibliography}{55}%
\makeatletter
\providecommand \@ifxundefined [1]{%
 \@ifx{#1\undefined}
}%
\providecommand \@ifnum [1]{%
 \ifnum #1\expandafter \@firstoftwo
 \else \expandafter \@secondoftwo
 \fi
}%
\providecommand \@ifx [1]{%
 \ifx #1\expandafter \@firstoftwo
 \else \expandafter \@secondoftwo
 \fi
}%
\providecommand \natexlab [1]{#1}%
\providecommand \enquote  [1]{``#1''}%
\providecommand \bibnamefont  [1]{#1}%
\providecommand \bibfnamefont [1]{#1}%
\providecommand \citenamefont [1]{#1}%
\providecommand \href@noop [0]{\@secondoftwo}%
\providecommand \href [0]{\begingroup \@sanitize@url \@href}%
\providecommand \@href[1]{\@@startlink{#1}\@@href}%
\providecommand \@@href[1]{\endgroup#1\@@endlink}%
\providecommand \@sanitize@url [0]{\catcode `\\12\catcode `\$12\catcode
  `\&12\catcode `\#12\catcode `\^12\catcode `\_12\catcode `\%12\relax}%
\providecommand \@@startlink[1]{}%
\providecommand \@@endlink[0]{}%
\providecommand \url  [0]{\begingroup\@sanitize@url \@url }%
\providecommand \@url [1]{\endgroup\@href {#1}{\urlprefix }}%
\providecommand \urlprefix  [0]{URL }%
\providecommand \Eprint [0]{\href }%
\providecommand \doibase [0]{http://dx.doi.org/}%
\providecommand \selectlanguage [0]{\@gobble}%
\providecommand \bibinfo  [0]{\@secondoftwo}%
\providecommand \bibfield  [0]{\@secondoftwo}%
\providecommand \translation [1]{[#1]}%
\providecommand \BibitemOpen [0]{}%
\providecommand \bibitemStop [0]{}%
\providecommand \bibitemNoStop [0]{.\EOS\space}%
\providecommand \EOS [0]{\spacefactor3000\relax}%
\providecommand \BibitemShut  [1]{\csname bibitem#1\endcsname}%
\let\auto@bib@innerbib\@empty
\bibitem [{\citenamefont {Miller}(2001)}]{Miller01}%
  \BibitemOpen
  \bibfield  {author} {\bibinfo {author} {\bibfnamefont {W.~H.}\ \bibnamefont
  {Miller}},\ }\href@noop {} {\bibfield  {journal} {\bibinfo  {journal} {J.
  Phys. Chem. A}\ }\textbf {\bibinfo {volume} {105}},\ \bibinfo {pages} {2942}
  (\bibinfo {year} {2001})}\BibitemShut {NoStop}%
\bibitem [{\citenamefont {Thoss}\ and\ \citenamefont {Wang}(2004)}]{Thoss04}%
  \BibitemOpen
  \bibfield  {author} {\bibinfo {author} {\bibfnamefont {M.}~\bibnamefont
  {Thoss}}\ and\ \bibinfo {author} {\bibfnamefont {H.}~\bibnamefont {Wang}},\
  }\href@noop {} {\bibfield  {journal} {\bibinfo  {journal} {Annu. Rev. Phys.
  Chem.}\ }\textbf {\bibinfo {volume} {55}},\ \bibinfo {pages} {299} (\bibinfo
  {year} {2004})}\BibitemShut {NoStop}%
\bibitem [{\citenamefont {Kay}(2005)}]{Kay05}%
  \BibitemOpen
  \bibfield  {author} {\bibinfo {author} {\bibfnamefont {K.~G.}\ \bibnamefont
  {Kay}},\ }\href@noop {} {\bibfield  {journal} {\bibinfo  {journal} {Annu.
  Rev. Phys. Chem.}\ }\textbf {\bibinfo {volume} {56}},\ \bibinfo {pages} {255}
  (\bibinfo {year} {2005})}\BibitemShut {NoStop}%
\bibitem [{\citenamefont {Miller}(2006)}]{Miller06}%
  \BibitemOpen
  \bibfield  {author} {\bibinfo {author} {\bibfnamefont {W.~H.}\ \bibnamefont
  {Miller}},\ }\href@noop {} {\bibfield  {journal} {\bibinfo  {journal} {J.
  Chem. Phys.}\ }\textbf {\bibinfo {volume} {125}},\ \bibinfo {pages} {132305}
  (\bibinfo {year} {2006})}\BibitemShut {NoStop}%
\bibitem [{\citenamefont {Koch}\ \emph {et~al.}(2008)\citenamefont {Koch},
  \citenamefont {Gro{\ss}mann}, \citenamefont {Stockburger},\ and\
  \citenamefont {Ankerhold}}]{Koch08}%
  \BibitemOpen
  \bibfield  {author} {\bibinfo {author} {\bibfnamefont {W.}~\bibnamefont
  {Koch}}, \bibinfo {author} {\bibfnamefont {F.}~\bibnamefont {Gro{\ss}mann}},
  \bibinfo {author} {\bibfnamefont {J.~T.}\ \bibnamefont {Stockburger}}, \ and\
  \bibinfo {author} {\bibfnamefont {J.}~\bibnamefont {Ankerhold}},\ }\href@noop
  {} {\bibfield  {journal} {\bibinfo  {journal} {Phys. Rev. Lett.}\ }\textbf
  {\bibinfo {volume} {100}},\ \bibinfo {pages} {230402} (\bibinfo {year}
  {2008})}\BibitemShut {NoStop}%
\bibitem [{\citenamefont {Moix}\ and\ \citenamefont {Pollak}(2008)}]{Moix08}%
  \BibitemOpen
  \bibfield  {author} {\bibinfo {author} {\bibfnamefont {J.~M.}\ \bibnamefont
  {Moix}}\ and\ \bibinfo {author} {\bibfnamefont {E.}~\bibnamefont {Pollak}},\
  }\href@noop {} {\bibfield  {journal} {\bibinfo  {journal} {J. Chem. Phys.}\
  }\textbf {\bibinfo {volume} {129}},\ \bibinfo {pages} {064515} (\bibinfo
  {year} {2008})}\BibitemShut {NoStop}%
\bibitem [{\citenamefont {Goletz}, \citenamefont {Koch},\ and\ \citenamefont
  {Gro{\ss}mann}(2010)}]{Goletz10}%
  \BibitemOpen
  \bibfield  {author} {\bibinfo {author} {\bibfnamefont {C.-M.}\ \bibnamefont
  {Goletz}}, \bibinfo {author} {\bibfnamefont {W.}~\bibnamefont {Koch}}, \ and\
  \bibinfo {author} {\bibfnamefont {F.}~\bibnamefont {Gro{\ss}mann}},\
  }\href@noop {} {\bibfield  {journal} {\bibinfo  {journal} {Chem. Phys.}\
  }\textbf {\bibinfo {volume} {375}},\ \bibinfo {pages} {227} (\bibinfo {year}
  {2010})}\BibitemShut {NoStop}%
\bibitem [{\citenamefont {{Van Vleck}}(1928)}]{VanVleck28}%
  \BibitemOpen
  \bibfield  {author} {\bibinfo {author} {\bibfnamefont {J.~H.}\ \bibnamefont
  {{Van Vleck}}},\ }\href@noop {} {\bibfield  {journal} {\bibinfo  {journal}
  {Proc. Natl. Acad. Sci.}\ }\textbf {\bibinfo {volume} {14}},\ \bibinfo
  {pages} {178} (\bibinfo {year} {1928})}\BibitemShut {NoStop}%
\bibitem [{\citenamefont {Klauder}\ and\ \citenamefont
  {Skagerstam}(1985)}]{Klauder85}%
  \BibitemOpen
  \bibfield  {author} {\bibinfo {author} {\bibfnamefont {J.~R.}\ \bibnamefont
  {Klauder}}\ and\ \bibinfo {author} {\bibfnamefont {B.-S.}\ \bibnamefont
  {Skagerstam}},\ }\href@noop {} {\emph {\bibinfo {title} {Coherent States:
  Applications in Physics and Mathematical Physics}}}\ (\bibinfo  {publisher}
  {World Scientific},\ \bibinfo {year} {1985})\BibitemShut {NoStop}%
\bibitem [{\citenamefont {Baranger}\ \emph {et~al.}(2001)\citenamefont
  {Baranger}, \citenamefont {de~Aguiar}, \citenamefont {Keck}, \citenamefont
  {Korsch},\ and\ \citenamefont {Schellhaa{\ss}}}]{Baranger01}%
  \BibitemOpen
  \bibfield  {author} {\bibinfo {author} {\bibfnamefont {M.}~\bibnamefont
  {Baranger}}, \bibinfo {author} {\bibfnamefont {M.~A.~M.}\ \bibnamefont
  {de~Aguiar}}, \bibinfo {author} {\bibfnamefont {F.}~\bibnamefont {Keck}},
  \bibinfo {author} {\bibfnamefont {H.~J.}\ \bibnamefont {Korsch}}, \ and\
  \bibinfo {author} {\bibfnamefont {B.}~\bibnamefont {Schellhaa{\ss}}},\
  }\href@noop {} {\bibfield  {journal} {\bibinfo  {journal} {J. Phys. A: Math.
  Gen.}\ }\textbf {\bibinfo {volume} {34}},\ \bibinfo {pages} {7227} (\bibinfo
  {year} {2001})}\BibitemShut {NoStop}%
\bibitem [{\citenamefont {Mart\'in-Fierro}\ and\ \citenamefont
  {Llorente}(2007)}]{Fierro07}%
  \BibitemOpen
  \bibfield  {author} {\bibinfo {author} {\bibfnamefont {E.}~\bibnamefont
  {Mart\'in-Fierro}}\ and\ \bibinfo {author} {\bibfnamefont {J.~M.~G.}\
  \bibnamefont {Llorente}},\ }\href@noop {} {\bibfield  {journal} {\bibinfo
  {journal} {J. Phys. A: Math. Gen.}\ }\textbf {\bibinfo {volume} {40}},\
  \bibinfo {pages} {1065} (\bibinfo {year} {2007})}\BibitemShut {NoStop}%
\bibitem [{\citenamefont {Braun}\ and\ \citenamefont {Garg}(2007)}]{Braun07}%
  \BibitemOpen
  \bibfield  {author} {\bibinfo {author} {\bibfnamefont {C.}~\bibnamefont
  {Braun}}\ and\ \bibinfo {author} {\bibfnamefont {A.}~\bibnamefont {Garg}},\
  }\href@noop {} {\bibfield  {journal} {\bibinfo  {journal} {J. Math. Phys.}\
  }\textbf {\bibinfo {volume} {48}},\ \bibinfo {pages} {032104} (\bibinfo
  {year} {2007})}\BibitemShut {NoStop}%
\bibitem [{\citenamefont {Huber}\ and\ \citenamefont {Heller}(1987)}]{Huber87}%
  \BibitemOpen
  \bibfield  {author} {\bibinfo {author} {\bibfnamefont {D.}~\bibnamefont
  {Huber}}\ and\ \bibinfo {author} {\bibfnamefont {E.~J.}\ \bibnamefont
  {Heller}},\ }\href@noop {} {\bibfield  {journal} {\bibinfo  {journal} {J.
  Chem. Phys.}\ }\textbf {\bibinfo {volume} {87}},\ \bibinfo {pages} {5302}
  (\bibinfo {year} {1987})}\BibitemShut {NoStop}%
\bibitem [{\citenamefont {Huber}, \citenamefont {Heller},\ and\ \citenamefont
  {Littlejohn}(1988)}]{Huber88}%
  \BibitemOpen
  \bibfield  {author} {\bibinfo {author} {\bibfnamefont {D.}~\bibnamefont
  {Huber}}, \bibinfo {author} {\bibfnamefont {E.~J.}\ \bibnamefont {Heller}}, \
  and\ \bibinfo {author} {\bibfnamefont {R.~G.}\ \bibnamefont {Littlejohn}},\
  }\href@noop {} {\bibfield  {journal} {\bibinfo  {journal} {J. Chem. Phys.}\
  }\textbf {\bibinfo {volume} {89}},\ \bibinfo {pages} {2003} (\bibinfo {year}
  {1988})}\BibitemShut {NoStop}%
\bibitem [{\citenamefont {Adachi}(1989)}]{Adachi89}%
  \BibitemOpen
  \bibfield  {author} {\bibinfo {author} {\bibfnamefont {S.}~\bibnamefont
  {Adachi}},\ }\href@noop {} {\bibfield  {journal} {\bibinfo  {journal} {Ann.
  Phys.}\ }\textbf {\bibinfo {volume} {195}},\ \bibinfo {pages} {45} (\bibinfo
  {year} {1989})}\BibitemShut {NoStop}%
\bibitem [{\citenamefont {Rubin}\ and\ \citenamefont
  {Klauder}(1995)}]{Rubin95}%
  \BibitemOpen
  \bibfield  {author} {\bibinfo {author} {\bibfnamefont {A.}~\bibnamefont
  {Rubin}}\ and\ \bibinfo {author} {\bibfnamefont {J.~R.}\ \bibnamefont
  {Klauder}},\ }\href@noop {} {\bibfield  {journal} {\bibinfo  {journal} {Ann.
  Phys.}\ }\textbf {\bibinfo {volume} {241}},\ \bibinfo {pages} {212} (\bibinfo
  {year} {1995})}\BibitemShut {NoStop}%
\bibitem [{\citenamefont {Shudo}\ and\ \citenamefont {Ikeda}(1995)}]{Shudo95}%
  \BibitemOpen
  \bibfield  {author} {\bibinfo {author} {\bibfnamefont {A.}~\bibnamefont
  {Shudo}}\ and\ \bibinfo {author} {\bibfnamefont {K.~S.}\ \bibnamefont
  {Ikeda}},\ }\href@noop {} {\bibfield  {journal} {\bibinfo  {journal} {Phys.
  Rev. Lett.}\ }\textbf {\bibinfo {volume} {74}},\ \bibinfo {pages} {682}
  (\bibinfo {year} {1995})}\BibitemShut {NoStop}%
\bibitem [{\citenamefont {Shudo}\ and\ \citenamefont {Ikeda}(1996)}]{Shudo96}%
  \BibitemOpen
  \bibfield  {author} {\bibinfo {author} {\bibfnamefont {A.}~\bibnamefont
  {Shudo}}\ and\ \bibinfo {author} {\bibfnamefont {K.~S.}\ \bibnamefont
  {Ikeda}},\ }\href@noop {} {\bibfield  {journal} {\bibinfo  {journal} {Phys.
  Rev. Lett.}\ }\textbf {\bibinfo {volume} {76}},\ \bibinfo {pages} {4151}
  (\bibinfo {year} {1996})}\BibitemShut {NoStop}%
\bibitem [{\citenamefont {Ribeiro}, \citenamefont {de~Aguiar},\ and\
  \citenamefont {Baranger}(2004)}]{Ribeiro04}%
  \BibitemOpen
  \bibfield  {author} {\bibinfo {author} {\bibfnamefont {A.~D.}\ \bibnamefont
  {Ribeiro}}, \bibinfo {author} {\bibfnamefont {M.~A.~M.}\ \bibnamefont
  {de~Aguiar}}, \ and\ \bibinfo {author} {\bibfnamefont {M.}~\bibnamefont
  {Baranger}},\ }\href@noop {} {\bibfield  {journal} {\bibinfo  {journal}
  {Phys. Rev. E}\ }\textbf {\bibinfo {volume} {69}},\ \bibinfo {pages} {066204}
  (\bibinfo {year} {2004})}\BibitemShut {NoStop}%
\bibitem [{\citenamefont {de~Aguiar}\ \emph {et~al.}(2005)\citenamefont
  {de~Aguiar}, \citenamefont {Baranger}, \citenamefont {Jaubert}, \citenamefont
  {Parisio},\ and\ \citenamefont {Ribeiro}}]{Aguiar05}%
  \BibitemOpen
  \bibfield  {author} {\bibinfo {author} {\bibfnamefont {M.~A.~M.}\
  \bibnamefont {de~Aguiar}}, \bibinfo {author} {\bibfnamefont {M.}~\bibnamefont
  {Baranger}}, \bibinfo {author} {\bibfnamefont {L.}~\bibnamefont {Jaubert}},
  \bibinfo {author} {\bibfnamefont {F.}~\bibnamefont {Parisio}}, \ and\
  \bibinfo {author} {\bibfnamefont {A.~D.}\ \bibnamefont {Ribeiro}},\
  }\href@noop {} {\bibfield  {journal} {\bibinfo  {journal} {J. Phys. A: Math.
  Gen.}\ }\textbf {\bibinfo {volume} {38}},\ \bibinfo {pages} {4645} (\bibinfo
  {year} {2005})}\BibitemShut {NoStop}%
\bibitem [{\citenamefont {Miller}(1970)}]{Miller70}%
  \BibitemOpen
  \bibfield  {author} {\bibinfo {author} {\bibfnamefont {W.~H.}\ \bibnamefont
  {Miller}},\ }\href@noop {} {\bibfield  {journal} {\bibinfo  {journal} {J.
  Chem. Phys.}\ }\textbf {\bibinfo {volume} {53}},\ \bibinfo {pages} {3578}
  (\bibinfo {year} {1970})}\BibitemShut {NoStop}%
\bibitem [{\citenamefont {Miller}(1974)}]{Miller74}%
  \BibitemOpen
  \bibfield  {author} {\bibinfo {author} {\bibfnamefont {W.~H.}\ \bibnamefont
  {Miller}},\ }\href@noop {} {\bibfield  {journal} {\bibinfo  {journal} {Adv.
  Chem. Phys.}\ }\textbf {\bibinfo {volume} {25}},\ \bibinfo {pages} {69}
  (\bibinfo {year} {1974})}\BibitemShut {NoStop}%
\bibitem [{\citenamefont {Heller}(1975)}]{Heller75}%
  \BibitemOpen
  \bibfield  {author} {\bibinfo {author} {\bibfnamefont {E.~J.}\ \bibnamefont
  {Heller}},\ }\href@noop {} {\bibfield  {journal} {\bibinfo  {journal} {J.
  Chem. Phys.}\ }\textbf {\bibinfo {volume} {62}},\ \bibinfo {pages} {1544}
  (\bibinfo {year} {1975})}\BibitemShut {NoStop}%
\bibitem [{\citenamefont {Herman}\ and\ \citenamefont {Kluk}(1984)}]{Herman84}%
  \BibitemOpen
  \bibfield  {author} {\bibinfo {author} {\bibfnamefont {M.~F.}\ \bibnamefont
  {Herman}}\ and\ \bibinfo {author} {\bibfnamefont {E.}~\bibnamefont {Kluk}},\
  }\href@noop {} {\bibfield  {journal} {\bibinfo  {journal} {Chem. Phys.}\
  }\textbf {\bibinfo {volume} {91}},\ \bibinfo {pages} {27} (\bibinfo {year}
  {1984})}\BibitemShut {NoStop}%
\bibitem [{\citenamefont {Kay}(1994{\natexlab{a}})}]{Kay94a}%
  \BibitemOpen
  \bibfield  {author} {\bibinfo {author} {\bibfnamefont {K.~G.}\ \bibnamefont
  {Kay}},\ }\href@noop {} {\bibfield  {journal} {\bibinfo  {journal} {J. Chem.
  Phys.}\ }\textbf {\bibinfo {volume} {100}},\ \bibinfo {pages} {4377}
  (\bibinfo {year} {1994}{\natexlab{a}})}\BibitemShut {NoStop}%
\bibitem [{\citenamefont {Kay}(1994{\natexlab{b}})}]{Kay94b}%
  \BibitemOpen
  \bibfield  {author} {\bibinfo {author} {\bibfnamefont {K.~G.}\ \bibnamefont
  {Kay}},\ }\href@noop {} {\bibfield  {journal} {\bibinfo  {journal} {J. Chem.
  Phys.}\ }\textbf {\bibinfo {volume} {100}},\ \bibinfo {pages} {4432}
  (\bibinfo {year} {1994}{\natexlab{b}})}\BibitemShut {NoStop}%
\bibitem [{\citenamefont {Kay}(1997)}]{Kay97}%
  \BibitemOpen
  \bibfield  {author} {\bibinfo {author} {\bibfnamefont {K.~G.}\ \bibnamefont
  {Kay}},\ }\href@noop {} {\bibfield  {journal} {\bibinfo  {journal} {J. Chem.
  Phys.}\ }\textbf {\bibinfo {volume} {107}},\ \bibinfo {pages} {2313}
  (\bibinfo {year} {1997})}\BibitemShut {NoStop}%
\bibitem [{\citenamefont {Zhang}\ and\ \citenamefont {Pollak}(2003)}]{Zhang03}%
  \BibitemOpen
  \bibfield  {author} {\bibinfo {author} {\bibfnamefont {S.}~\bibnamefont
  {Zhang}}\ and\ \bibinfo {author} {\bibfnamefont {E.}~\bibnamefont {Pollak}},\
  }\href@noop {} {\bibfield  {journal} {\bibinfo  {journal} {Phys. Rev. Lett.}\
  }\textbf {\bibinfo {volume} {91}},\ \bibinfo {pages} {190201} (\bibinfo
  {year} {2003})}\BibitemShut {NoStop}%
\bibitem [{\citenamefont {Zhang}\ and\ \citenamefont {Pollak}(2004)}]{Zhang04}%
  \BibitemOpen
  \bibfield  {author} {\bibinfo {author} {\bibfnamefont {D.~H.}\ \bibnamefont
  {Zhang}}\ and\ \bibinfo {author} {\bibfnamefont {E.}~\bibnamefont {Pollak}},\
  }\href@noop {} {\bibfield  {journal} {\bibinfo  {journal} {Phys. Rev. Lett.}\
  }\textbf {\bibinfo {volume} {93}},\ \bibinfo {pages} {140401} (\bibinfo
  {year} {2004})}\BibitemShut {NoStop}%
\bibitem [{\citenamefont {Heller}(1991)}]{Heller91}%
  \BibitemOpen
  \bibfield  {author} {\bibinfo {author} {\bibfnamefont {E.~J.}\ \bibnamefont
  {Heller}},\ }\href@noop {} {\bibfield  {journal} {\bibinfo  {journal} {J.
  Chem. Phys.}\ }\textbf {\bibinfo {volume} {94}},\ \bibinfo {pages} {2723}
  (\bibinfo {year} {1991})}\BibitemShut {NoStop}%
\bibitem [{\citenamefont {Tomsovic}\ and\ \citenamefont
  {Heller}(1991)}]{Tomsovic91}%
  \BibitemOpen
  \bibfield  {author} {\bibinfo {author} {\bibfnamefont {S.}~\bibnamefont
  {Tomsovic}}\ and\ \bibinfo {author} {\bibfnamefont {E.}~\bibnamefont
  {Heller}},\ }\href@noop {} {\bibfield  {journal} {\bibinfo  {journal} {Phys.
  Rev. Lett.}\ }\textbf {\bibinfo {volume} {67}},\ \bibinfo {pages} {664}
  (\bibinfo {year} {1991})}\BibitemShut {NoStop}%
\bibitem [{\citenamefont {Shalashilin}\ and\ \citenamefont
  {Child}(2004)}]{Shalashilin04}%
  \BibitemOpen
  \bibfield  {author} {\bibinfo {author} {\bibfnamefont {D.~V.}\ \bibnamefont
  {Shalashilin}}\ and\ \bibinfo {author} {\bibfnamefont {M.~S.}\ \bibnamefont
  {Child}},\ }\href@noop {} {\bibfield  {journal} {\bibinfo  {journal} {Chem.
  Phys.}\ }\textbf {\bibinfo {volume} {304}},\ \bibinfo {pages} {103} (\bibinfo
  {year} {2004})}\BibitemShut {NoStop}%
\bibitem [{\citenamefont {Shalashilin}\ and\ \citenamefont
  {Burghardt}(2008)}]{Shalashilin08}%
  \BibitemOpen
  \bibfield  {author} {\bibinfo {author} {\bibfnamefont {D.~V.}\ \bibnamefont
  {Shalashilin}}\ and\ \bibinfo {author} {\bibfnamefont {I.}~\bibnamefont
  {Burghardt}},\ }\href@noop {} {\bibfield  {journal} {\bibinfo  {journal} {J.
  Chem. Phys.}\ }\textbf {\bibinfo {volume} {129}},\ \bibinfo {pages} {084104}
  (\bibinfo {year} {2008})}\BibitemShut {NoStop}%
\bibitem [{\citenamefont {Pollak}\ and\ \citenamefont {Shao}(2003)}]{Pollak03}%
  \BibitemOpen
  \bibfield  {author} {\bibinfo {author} {\bibfnamefont {E.}~\bibnamefont
  {Pollak}}\ and\ \bibinfo {author} {\bibfnamefont {J.}~\bibnamefont {Shao}},\
  }\href@noop {} {\bibfield  {journal} {\bibinfo  {journal} {J. Phys. Chem. A}\
  }\textbf {\bibinfo {volume} {107}},\ \bibinfo {pages} {7112} (\bibinfo {year}
  {2003})}\BibitemShut {NoStop}%
\bibitem [{\citenamefont {Kay}(2006)}]{Kay06}%
  \BibitemOpen
  \bibfield  {author} {\bibinfo {author} {\bibfnamefont {K.~G.}\ \bibnamefont
  {Kay}},\ }\href@noop {} {\bibfield  {journal} {\bibinfo  {journal} {Chem.
  Phys.}\ }\textbf {\bibinfo {volume} {322}},\ \bibinfo {pages} {3} (\bibinfo
  {year} {2006})}\BibitemShut {NoStop}%
\bibitem [{\citenamefont {de~Aguiar}, \citenamefont {Vitiello},\ and\
  \citenamefont {Grigolo}(2010)}]{Aguiar10}%
  \BibitemOpen
  \bibfield  {author} {\bibinfo {author} {\bibfnamefont {M.~A.~M.}\
  \bibnamefont {de~Aguiar}}, \bibinfo {author} {\bibfnamefont {S.~A.}\
  \bibnamefont {Vitiello}}, \ and\ \bibinfo {author} {\bibfnamefont
  {A.}~\bibnamefont {Grigolo}},\ }\href@noop {} {\bibfield  {journal} {\bibinfo
   {journal} {Chem. Phys.}\ }\textbf {\bibinfo {volume} {370}},\ \bibinfo
  {pages} {42} (\bibinfo {year} {2010})}\BibitemShut {NoStop}%
\bibitem [{\citenamefont {Viscondi}\ and\ \citenamefont
  {de~Aguiar}()}]{Viscondi11b}%
  \BibitemOpen
  \bibfield  {author} {\bibinfo {author} {\bibfnamefont {T.~F.}\ \bibnamefont
  {Viscondi}}\ and\ \bibinfo {author} {\bibfnamefont {M.~A.~M.}\ \bibnamefont
  {de~Aguiar}},\ }\href@noop {} {}\bibinfo {note} {{arXiv:1103.0958v1
  [math-ph]}}\BibitemShut {NoStop}%
\bibitem [{\citenamefont {Gilmore}, \citenamefont {Bowden},\ and\ \citenamefont
  {Narducci}(1975)}]{Gilmore75}%
  \BibitemOpen
  \bibfield  {author} {\bibinfo {author} {\bibfnamefont {R.}~\bibnamefont
  {Gilmore}}, \bibinfo {author} {\bibfnamefont {C.~M.}\ \bibnamefont {Bowden}},
  \ and\ \bibinfo {author} {\bibfnamefont {L.~M.}\ \bibnamefont {Narducci}},\
  }\href@noop {} {\bibfield  {journal} {\bibinfo  {journal} {Phys. Rev. A}\
  }\textbf {\bibinfo {volume} {12}},\ \bibinfo {pages} {1019} (\bibinfo {year}
  {1975})}\BibitemShut {NoStop}%
\bibitem [{Note1()}]{Note1}%
  \BibitemOpen
  \bibinfo {note} {According to the adopted notation, the juxtaposition of two
  vectors $a$ and $b$ represents the matrix product
  $ab=a_{1}b_{1}+a_{2}b_{2}+\protect \ldots +a_{n-1}b_{n-1}$.}\BibitemShut
  {Stop}%
\bibitem [{Note2()}]{Note2}%
  \BibitemOpen
  \bibinfo {note} {For simplicity, in what follows we choose the system of
  units so that $\hbar =1$.}\BibitemShut {Stop}%
\bibitem [{Note3()}]{Note3}%
  \BibitemOpen
  \bibinfo {note} {Considering two vector quantities $a$ and $b$, we denote by
  $\protect \frac {\partial a}{\partial b}$ the matrix whose elements follow
  from $\left [\protect \frac {\partial a}{\partial b}\right ]_{jk}=\protect
  \frac {\partial a_{j}}{\partial b_{k}}$, with $j,k=1,2,\protect \ldots
  ,(n-1)$. In the case of a scalar function $f(a)$, we have that $\protect
  \frac {\partial f(a)}{\partial a}$ represents a vector whose entries are
  given by $\left [\protect \frac {\partial f(a)}{\partial a}\right
  ]_{j}=\protect \frac {\partial f(a)}{\partial a_{j}}$, also for
  $j=1,2,\protect \ldots ,(n-1)$.}\BibitemShut {Stop}%
\bibitem [{Note4()}]{Note4}%
  \BibitemOpen
  \bibinfo {note} {In the equation \protect \textup {\hbox {\mathsurround \z@
  \protect \normalfont (\ignorespaces \ref {eq1p6}\unskip \@@italiccorr )}} we
  introduce the notation for the dyadic product. That is, considering two
  arbitrary vectors $a$ and $b$ of dimension $(n-1)$, the outcome of the
  product $a\otimes b$ is a matrix with elements given by $(a\otimes
  b)_{jk}=a_{j}b_{k}$.}\BibitemShut {Stop}%
\bibitem [{Note5()}]{Note5}%
  \BibitemOpen
  \bibinfo {note} {Due to the overcompleteness of the coherent states, there
  are several ways to perform the semiclassical approximation of the
  propagator, resulting from different quantization schemes (choices of
  operator ordering). Each one of these corresponds to a distinct correction
  term\cite {Baranger01, Santos06}.}\BibitemShut {Stop}%
\bibitem [{Note6()}]{Note6}%
  \BibitemOpen
  \bibinfo {note} {A focal point represents a crossing between trajectories
  when projected onto a particular subspace of the complete phase
  space.}\BibitemShut {Stop}%
\bibitem [{Note7()}]{Note7}%
  \BibitemOpen
  \bibinfo {note} {In fact, as we shall see below, the focal points correspond
  to zeros of the whole integrand in the initial value
  representation.}\BibitemShut {Stop}%
\bibitem [{Note8()}]{Note8}%
  \BibitemOpen
  \bibinfo {note} {Note that, for simplicity of notation, we omit the subindex
  `$f$' for the final condition of the semiclassical propagator. In this way we
  also emphasize the role of the variables $w$ as coordinates of a classical
  phase space in which we can represent the quantum states and
  operators.}\BibitemShut {Stop}%
\bibitem [{\citenamefont {Scully}\ and\ \citenamefont
  {Zubairy}(1997)}]{Scully97}%
  \BibitemOpen
  \bibfield  {author} {\bibinfo {author} {\bibfnamefont {M.~O.}\ \bibnamefont
  {Scully}}\ and\ \bibinfo {author} {\bibfnamefont {M.~S.}\ \bibnamefont
  {Zubairy}},\ }\href@noop {} {\emph {\bibinfo {title} {Quantum Optics}}}\
  (\bibinfo  {publisher} {Cambridge University Press},\ \bibinfo {year}
  {1997})\BibitemShut {NoStop}%
\bibitem [{\citenamefont {Viscondi}\ and\ \citenamefont
  {Furuya}()}]{Viscondi11a}%
  \BibitemOpen
  \bibfield  {author} {\bibinfo {author} {\bibfnamefont {T.~F.}\ \bibnamefont
  {Viscondi}}\ and\ \bibinfo {author} {\bibfnamefont {K.}~\bibnamefont
  {Furuya}},\ }\href@noop {} {}\bibinfo {note} {{arXiv:1011.1138v1
  [quant-ph]}}\BibitemShut {NoStop}%
\bibitem [{\citenamefont {Negele}\ and\ \citenamefont
  {Orland}(1998)}]{Negele98}%
  \BibitemOpen
  \bibfield  {author} {\bibinfo {author} {\bibfnamefont {J.~W.}\ \bibnamefont
  {Negele}}\ and\ \bibinfo {author} {\bibfnamefont {H.}~\bibnamefont
  {Orland}},\ }\href@noop {} {\emph {\bibinfo {title} {Quantum Many-Particle
  Systems}}}\ (\bibinfo  {publisher} {Westview Press},\ \bibinfo {year}
  {1998})\BibitemShut {NoStop}%
\bibitem [{Note9()}]{Note9}%
  \BibitemOpen
  \bibinfo {note} {Notice that the action $S$ and the correction term $I$ are
  real valued when calculated on the principal trajectory. This property makes
  the removal of the principal trajectory by the heuristic filter a very
  unlikely event, as can be inferred from the discussion below the inequality
  \protect \textup {\hbox {\mathsurround \z@ \protect \normalfont
  (\ignorespaces \ref {eq1p22}\unskip \@@italiccorr )}}.}\BibitemShut {Stop}%
\bibitem [{\citenamefont {Zhang}, \citenamefont {Feng},\ and\ \citenamefont
  {Gilmore}(1990)}]{Zhang90}%
  \BibitemOpen
  \bibfield  {author} {\bibinfo {author} {\bibfnamefont {W.-M.}\ \bibnamefont
  {Zhang}}, \bibinfo {author} {\bibfnamefont {D.~H.}\ \bibnamefont {Feng}}, \
  and\ \bibinfo {author} {\bibfnamefont {R.}~\bibnamefont {Gilmore}},\
  }\href@noop {} {\bibfield  {journal} {\bibinfo  {journal} {Rev. Mod. Phys.}\
  }\textbf {\bibinfo {volume} {62}},\ \bibinfo {pages} {867} (\bibinfo {year}
  {1990})}\BibitemShut {NoStop}%
\bibitem [{\citenamefont {Yaffe}(1982)}]{Yaffe82}%
  \BibitemOpen
  \bibfield  {author} {\bibinfo {author} {\bibfnamefont {L.~G.}\ \bibnamefont
  {Yaffe}},\ }\href@noop {} {\bibfield  {journal} {\bibinfo  {journal} {Rev.
  Mod. Phys.}\ }\textbf {\bibinfo {volume} {54}},\ \bibinfo {pages} {407}
  (\bibinfo {year} {1982})}\BibitemShut {NoStop}%
\bibitem [{\citenamefont {Viscondi}, \citenamefont {Furuya},\ and\
  \citenamefont {de~Oliveira}(2010)}]{Viscondi10}%
  \BibitemOpen
  \bibfield  {author} {\bibinfo {author} {\bibfnamefont {T.~F.}\ \bibnamefont
  {Viscondi}}, \bibinfo {author} {\bibfnamefont {K.}~\bibnamefont {Furuya}}, \
  and\ \bibinfo {author} {\bibfnamefont {M.~C.}\ \bibnamefont {de~Oliveira}},\
  }\href@noop {} {\bibfield  {journal} {\bibinfo  {journal} {EPL}\ }\textbf
  {\bibinfo {volume} {90}},\ \bibinfo {pages} {10014} (\bibinfo {year}
  {2010})}\BibitemShut {NoStop}%
\bibitem [{\citenamefont {Arecchi}\ \emph {et~al.}(1972)\citenamefont
  {Arecchi}, \citenamefont {Courtens}, \citenamefont {Gilmore},\ and\
  \citenamefont {Thomas}}]{Arecchi72}%
  \BibitemOpen
  \bibfield  {author} {\bibinfo {author} {\bibfnamefont {F.~T.}\ \bibnamefont
  {Arecchi}}, \bibinfo {author} {\bibfnamefont {E.}~\bibnamefont {Courtens}},
  \bibinfo {author} {\bibfnamefont {R.}~\bibnamefont {Gilmore}}, \ and\
  \bibinfo {author} {\bibfnamefont {H.}~\bibnamefont {Thomas}},\ }\href@noop {}
  {\bibfield  {journal} {\bibinfo  {journal} {Phys. Rev. A}\ }\textbf {\bibinfo
  {volume} {6}},\ \bibinfo {pages} {2211} (\bibinfo {year} {1972})}\BibitemShut
  {NoStop}%
\bibitem [{\citenamefont {dos Santos}\ and\ \citenamefont
  {de~Aguiar}(2006)}]{Santos06}%
  \BibitemOpen
  \bibfield  {author} {\bibinfo {author} {\bibfnamefont {L.~C.}\ \bibnamefont
  {dos Santos}}\ and\ \bibinfo {author} {\bibfnamefont {M.~A.~M.}\ \bibnamefont
  {de~Aguiar}},\ }\href@noop {} {\bibfield  {journal} {\bibinfo  {journal} {J.
  Phys. A: Math. Gen.}\ }\textbf {\bibinfo {volume} {39}},\ \bibinfo {pages}
  {13465} (\bibinfo {year} {2006})}\BibitemShut {NoStop}%
\end{thebibliography}%

\end{document}